\documentclass[aps,pra,preprint,superscriptaddress,longbibliography]{revtex4-1}

\usepackage{amsmath}
\usepackage{amsthm}
\usepackage{amsfonts}
\usepackage{amssymb}
\usepackage{xcolor,graphicx}
\usepackage{tikz}
\usepackage{color}
\usepackage[pdftex]{hyperref}
\usepackage{longtable}
\usepackage{array}
\usepackage{bigints}
\usepackage{dsfont}
\begin{document}
\bibliographystyle{apsrev4-1}

\setkeys{Gin}{draft=false} 

\title{Revisiting Einstein's analogy: black holes as gradient-index lenses}

\author{J. E. G\'{o}mez-Correa}
\email{jgomez@inaoep.mx}
\affiliation{Instituto Nacional de Astrof\'isica, \'Optica y Electr\'onica, Santa María Tonantzintla, Puebla, 72840, Mexico}
\author{E. Esp\'{i}ndola-Ramos}
\email{ernestoespindola@inaoep.mx}
\affiliation{Instituto Nacional de Astrof\'isica, \'Optica y Electr\'onica, Santa María Tonantzintla, Puebla, 72840, Mexico}

\author{O. L\'{o}pez-Cruz}
\email{omarlx@inaoep.mx}
\affiliation{Instituto Nacional de Astrof\'isica, \'Optica y Electr\'onica, Santa María Tonantzintla, Puebla, 72840, Mexico}
\author{S. Ch\'{a}vez-Cerda}
\email{sabino@inaoep.mx}
\affiliation{Instituto Nacional de Astrof\'isica, \'Optica y Electr\'onica, Santa María Tonantzintla, Puebla, 72840, Mexico}


\begin{abstract}
According to Albert Einstein, gravitation is analogous to an optical medium. Building on this idea, various definitions of the gradient-index (GRIN) medium representing curved spacetime have been proposed; often, these approaches demand advanced knowledge of General Relativity (GR) and its associated mathematical methods. 
This paper introduces an alternative approach for generating the form of GRIN media that reproduces the exact behavior of photon trajectories in the presence of a Schwarzschild black hole or about the equatorial plane of a Kerr black hole. Our approach is based on Noether's theorem that leads to the optical Fermat principle, leveraging the system's symmetry, \textcolor{black}{resulting in an unrestricted} \textcolor{black}{and robust} method for finding GRIN media in a flat spacetime that reproduces the behavior of null geodesics in Schwarzschild-like spacetimes,  {\textit{employing only photon trajectories as input}}. This approach is also valid for massive particles and provides a complementary understanding of GR. \textcolor{black}{For this reason, the method is applied to generate the GRIN medium from the S2 star's trajectory around SgrA*, demonstrating that sufficient precision can be achieved to reproduce the orbit's subtle apsidal precession.}
\end{abstract}
\maketitle




\section{Introduction}
\label{sect:intro}  
\textcolor{black}{Gravitational lensing is the phenomenon in which waves (including gravitational waves) or particles are deflected due to the curvature of spacetime}. \textcolor{black}{The idea that gravity affects light has been considered for a long time. In the first edition of Opticks (1704), Isaac Newton suggested that bodies could influence the behavior of light, implying a deflection of its rays \cite{Schneider_1992}. In 1801, Johann Georg von Soldner published a quantitative analysis of this effect within the framework of Newtonian mechanics. Independently, Albert Einstein first described the effect of gravitation {\color{black}on light} in 1911 within the context of special relativity {\color{black}\cite{Einstein_1911}}. Four years later, he extended his analysis to the general theory of relativity (GR), thus incorporating the curvature of spacetime and providing a precise quantitative prediction of light deflection {\color{black}\cite{einstein1915perihelion}}. Since then, gravitational lensing has been extensively studied through various spacetime solutions to Einstein's field equations, {\color{black}proving to be a powerful tool for exploring the universe}}.

Many astrophysical studies rely on strong, weak, and microlensing phenomena to understand galaxy clusters' structure and mass \textcolor{black}{distributions}. As natural cosmic lenses, these phenomena have led to the discovery of exoplanets beyond our solar system and have the potential to extend our observational capabilities beyond light. With advancements in gravitational wave detectors, the possibility of detecting lensed gravitational waves is increasing. For a comprehensive review of these subjects, \textcolor{black}{we refer the interested reader to} \cite{Perlick_2004, Rodr_guez_Fajardo_2023, Bartelmann_2010, Grespan_2023}.

The deflection of light is often described using the thin-lens approximation. However, the direct detection of gravitational waves by the Laser Interferometer Gravitational-Wave Observatory (LIGO) \cite{Raab_2017} and the observation of the \textcolor{black}{black hole shadow} at the center of our galaxy by the Event Horizon Telescope collaboration (EHT) \cite{Castelvecchi_2019} promote the use of the complete theory of GR. Efforts in this direction include the Virbhadra-Ellis lens equation model \cite{Virbhadra_2000}, which is valid for large deflection angles of rays, \textcolor{black}{and the description of other effects \cite{Virbhadra_2009,Virbhadra_2022,Virbhadra_2024,Virbhadra_2024_c}.} 

{\color{black}Although the black hole shadow is already a manifestation of strong gravitational fields, next-generation telescopes such as the ngEHT \cite{Johnson_2023} will improve our ability to observe additional relativistic features that remain unresolved with current instrumentation. These include the photon ring, the inner shadow, and dynamic signatures such as light echoes and polarization structures near the event horizon.} {\color{black}For a historical overview of gravitational lensing, the reader is referred to the foundational works already mentioned above. For comprehensive reviews of gravitational lensing theory and its applications in astrophysics and cosmology, see \cite{Perlick_2004, Bartelmann_2010}. For more recent studies on strong lensing by Schwarzschild black holes, including image distortions and relativistic images, see \cite{Virbhadra_2000,Virbhadra_2009,Virbhadra_2022,Virbhadra_2024,Virbhadra_2024_c}.}

Although phenomena involving the strong effects of curved spacetime are beyond our capacity to directly experience or experiment with, analogies with other branches of physics, such as classical mechanics, quantum mechanics, optics, and fluid mechanics \cite{Plebanski_1960, de_Felice_1971,Sch_tzhold_2002,Philbin_2008,Genov_2009,Chen_2010,leonhardt2010geometry,Tinguely_2020,Rodr_guez_Fajardo_2023,_van_ara_2024}, may enable their analogue simulations in the laboratory.

Approaches to studying light propagation in curved spacetimes using concepts from optics and variational principles have been explored since the birth of GR \textcolor{black}{\cite{eddington1920report,eddington1921space}}. As noted in Ref.~\cite{de_Felice_1971}, ``\emph{the idea that gravitation is equivalent to an optical medium was first suggested by Einstein himself}.'' Key concepts along this path include Maxwell's equations in GR as electrodynamics in a macroscopic medium, the optical metric, and a relativistic version of Fermat's principle , among others \cite{Plebanski_1960, de_Felice_1971,Gordon_1923,Kovner_1990}.

Metamaterials allow the reproduction of the behavior of the electromagnetic waves in curved spacetimes through the formal equivalence between Maxwell's equations in GR and those of electrodynamics in a macroscopic medium. However, reproducing light rays through GRIN media is advantageous when polarization effects are neglected, as it is much simpler and easier to construct experimentally \cite{Tinguely_2020}. Currently, various definitions of the gradient index for curved \textcolor{black}{spacetimes} that mimics photon trajectories exist; therefore, following the comments in Refs.~\cite{Parvizi_2024, Ramezani_Aval_2024}, the core of the matter is how to design optical devices or metamaterials that exactly reproduce photon trajectories.

Noether's theorem, one of the most fascinating theorems in modern physics, was inspired by Albert Einstein's work on relativity. It states that a natural law described by a Lagrangian function that exhibits symmetry \textcolor{black}{has an invariant, a conserved quantity.} The associated Euler–Lagrange equations inherit the symmetry group of the variational problem \cite{kogan2003invariant}. In general, variational principles are related to finding extrema that, in physical systems, are commonly minima. In this sense, Fermat's principle states that a ray of light traveling in a medium with variable refractive index will take the route that minimizes the time of travel between two given points, which can be expressed as a variational principle yielding to the corresponding Euler-Lagrange equations, \textcolor{black}{substantially simplifying} the problem.

Suppose the \textcolor{black}{geometrical optics} formalism is used alone. In that case, finding the path of a ray of light in Gradient-Index (GRIN) media is cumbersome since it requires finding the solution of a partial differential equation called the Eikonal. Different methods have been proposed to solve this equation, but their implementation is complicated and is restricted to a medium with a given GRIN distribution. 

In recent decades, the inverse problem—finding the GRIN distribution for a \textcolor{black}{given light path}—has garnered increasing attention. Since this task may be even more challenging than determining ray propagation in a GRIN medium, the methods proposed in the literature \textcolor{black}{have different complexities}. However, it has been demonstrated that an exact method for solving the inverse problem in any symmetric GRIN medium can be derived using Noether's invariant of the system \cite{G_mez_Correa_2023}. This approach, known as the Physical GRIN Reconstruction (PhysGRIN) method, enables the determination of the refractive index at any point within the medium based on the ray direction and Noether's invariant. Following the ray's path and employing the invariant effectively defines the symmetric refractive index distribution. Once the ray trajectories are established, the algorithm is easily implemented and computationally efficient. This method, being formulated using the variational principle, can be extended to other areas of physics, such as astronomy \cite{Schneider_1992,Price_2018,thorne2000gravitation,landau1960mechanics,wambsganss1998gravitational}.

In this paper, using the PhysGRIN method and leveraging the trajectory of a photon in the equatorial plane of a Schwarzschild black hole \textcolor{black}{or in the equatorial plane} of a Kerr black hole, we introduce an \textcolor{black}{alternative} approach for constructing GRIN lenses that replicate photon paths in the curved spacetime of these black holes. Inspired by Einstein's analogy between gravitation and optical media, this method allows us to create GRIN \textcolor{black}{models of curved spacetimes} without requiring advanced knowledge of GR or tensor calculus, instead relying on Noether's theorem and the system's symmetry. \textcolor{black}{ The robustness of the PhysGRIN method is tested by adding noise to exact null geodesics, followed by a discussion on how to best handle stochastic data. Since PhysGRIN relies solely on a given trajectory, it can also be applied to massive particles. As a case study, the gradient index profile for the orbit of the S2 star around SgrA*, the nearest candidate supermassive black hole, is reconstructed, demonstrating that the method is both simple and accurate enough to reproduce its subtle apsidal precession \cite{2020}.}

\section{Photon trajectories in the equatorial plane of Kerr black holes}
 \label{photon_trajectories}

The Schwarzschild metric is one of the most studied models of black holes; it is the unique, static, spherically symmetric vacuum solution to Einstein's field equations and provides a fundamental framework for investigating \textcolor{black}{gravitational lensing} (see \cite{Schneider_1992,Virbhadra_2009}, and references therein). It is also a particular case of the Kerr spacetime, which generalizes it to include the effects of rotation.

\textcolor{black}{The line element} for a Kerr black hole of mass $M$ in Boyer–Lindquist coordinates is given by
\begin{eqnarray}
 \nonumber
\textcolor{black}{\text{d}}s^{2}= - c^2 \left( 1-\frac{2GM r_{bl}}{c^{2} \Sigma} \right)  \textcolor{black}{\text{d}}t^2  -\frac{4GMr_{bl}a\sin^2\theta }{c^2\Sigma}c\textcolor{black}{\text{d}}t\textcolor{black}{\mathrm{d}}\phi  \\ \label{Kerr}
 + \left( r_{bl}^2 + a^2 +   \frac{2GMr_{bl}a^2}{c^2\Sigma} \sin^2 \theta   \right)\sin^2\theta \textcolor{black}{\mathrm{d}}\phi^2  
 + \frac{\Sigma}{\Delta} \textcolor{black}{\mathrm{d}}r_{bl}^2 + \Sigma \textcolor{black}{\mathrm{d}}\theta^2,
\end{eqnarray}
where $G$ denotes the universal gravitational constant, $c$ the speed of light, $a$ is related to the angular momentum $J$ of the black hole by $ a = J/Mc $, and
\begin{eqnarray}
    \Sigma = r_{bl}^2 + a^2\cos^2\theta,\quad \Delta = r_{bl}^2-\frac{2GMr_{bl}}{c^2}+a^2.
\end{eqnarray}
This metric has two coordinate singularities determined by $\Delta = 0$. The exterior singularity, $r_{bl,+} = \left[ GM + \sqrt{G^2 M^2 - c^4 a^2} \right] / c^2$, is known as the outer event horizon and marks the boundary beyond where nothing can escape. The interior singularity, $r_{bl,-} = \left[ GM - \sqrt{G^2 M^2 - c^4 a^2} \right] / c^2$, is called the inner event horizon or Cauchy horizon, and marks the region where \textcolor{black}{the spacetime} becomes unstable \cite{mcnamara1978instability}. These coordinate singularities restrict the black hole's angular momentum, which must satisfy $-GM/c \leq a \leq GM/c$ for $r_{bl,+}$ and $r_{bl,-}$ to be real. 

The Kerr metric also contains a true singularity at $r_{bl} = 0$ and $ \theta = \pi/2$, where $\Sigma = 0$, causing the curvature to become infinite. \textcolor{black}{It is a common depiction to represent such a singularity as a ring of radius $a$ in the equatorial plane. This follows from the relationship between Boyer-Lindquist and Cartesian coordinates: $ x_{c} = \sqrt{r_{bl}^2+a^2}\sin\theta\cos\phi $, $ y_{c} = \sqrt{r_{bl}^2+a^2}\sin\theta\sin\phi $, and $ z_{c} = r_{bl}\cos\theta$. For this reason, from now on, we will visualize the null geodesics graphically in Cartesian coordinates.}  

Unlike the Schwarzschild metric $(a = 0)$, the coefficient of $\textcolor{black}{\mathrm{d}}t^2$ in the Kerr metric changes sign at a region different from the event horizon, known as the \textcolor{black}{ergoregion}. The boundary of the \textcolor{black}{ergoregion (ergosurface)} is given by $r_{bl,e} = \left[ GM + \sqrt{G^2 M^2 - a^2 c^4 \cos^2 \theta} \right] / c^2$, which is exterior to the event horizon except at the poles $(\theta = 0, \pi)$, where they coincide. Inside the \textcolor{black}{ergoregion}, static particles $(\textcolor{black}{\mathrm{d}}r_{bl} = \textcolor{black}{\mathrm{d}}\theta = \textcolor{black}{\mathrm{d}}\phi = 0)$ 
\textcolor{black}{cannot exist} because their worldlines would be forced to move in the direction of the black hole's angular momentum due to the \textcolor{black}{frame-dragging effect.}

Eq. (\ref{Kerr}) can be summarized using the metric tensor notation as follows:
\begin{eqnarray}
\label{metric_tensor}
    \textcolor{black}{\mathrm{d}}s^{2}=g_{\alpha \beta}\textcolor{black}{\mathrm{d}}x^{\alpha}\textcolor{black}{\mathrm{d}}x^{\beta},
\end{eqnarray}
where the upper and lower indices denote Einstein's summation notation, with $\alpha, \beta = 0, 1, 2, 3$, and $x^{0} = ct$, $x^{1} = r$, $x^{2} = \theta$, and $x^{3} = \phi$. The world line of a particle, which represents the path that the particle traces in spacetime, is a one-dimensional curve in four-dimensional spacetime. {\color{black}This curve can be parameterized by a parameter $\lambda$, which labels each point along the trajectory. For massive particles, it is convenient to identify $\lambda$ with the particle's proper time $\tau$. However, since proper time is not defined for massless particles such as photons, $\lambda$ is instead interpreted as an affine parameter.} The parameterization of the world line is denoted as
\begin{eqnarray}
    {\bf x} = ({x^{0}(\lambda),x^{1}(\lambda),x^{2}(\lambda),x^{3}(\lambda)}).
\end{eqnarray}
The standard approach to null geodesics is through the Lagrangian formalism, though they can also be expressed in the Hamiltonian formalism. A brief discussion on these formalisms and the challenges of computing particle or photon trajectories can be found in Ref.~\cite{Price_2018}. Since we are interested in geodesics confined to the equatorial plane to exploit spherical symmetry for GRIN medium reconstruction, the most straightforward approach is via \textcolor{black}{the Lagrangian} formalism. The Lagrangian associated with the metric (\ref{metric_tensor}) is given by:
\begin{eqnarray}
\label{lagrangian}
    \mathcal{L}=\frac{1}{2} g_{\alpha\beta} \dot{x}^{\alpha} \dot{x}^{\beta},
\end{eqnarray}
where $\dot{x}^{\alpha} = \textcolor{black}{\text{d}}x^{\alpha} / \textcolor{black}{\mathrm{d}}\lambda$ denotes the components of the four-velocity. The term $g(\dot{x},\dot{x}) \equiv g_{\alpha\beta} \dot{x}^{\alpha} \dot{x}^{\beta}$ represents the square of the norm of the four-velocity, which for photons equals zero, as the four-velocity vector lies on the local light cone. Under these conditions,  null geodesics (world lines of photons) satisfy the following system of equations:
\begin{eqnarray}
\label{Euler_Lagrange}
    \mathcal{L} = 0, \quad \frac{\textcolor{black}{\mathrm{d}}}{\textcolor{black}{\mathrm{d}}\lambda} \frac{\partial \mathcal{L}}{\partial \dot{x}^{\alpha}} - \frac{\partial \mathcal{L}}{\partial x^{\alpha}} = 0.
\end{eqnarray}
A review and classification of trajectories using elliptic integrals of the first and third kind, Carlson's elliptic integrals, Jacobi-elliptic functions, and Weierstrass elliptic functions, along with codes for computing geodesics in Wolfram Mathematica and Fortran, can be found in Ref.~\cite{Dexter_2009,Yang_2013,fan2018analytical,Gralla_2020,Cie_lik_2023,Liu__2024}.

\textcolor{black}{ At this point, it is important to clarify that this work aims to demonstrate the simplicity and accuracy of PhysGRIN in reconstructing a GRIN medium that reproduces photon trajectories, which can be exploited without major complications when light trajectories exhibit symmetries. The method used to calculate the null geodesics (whether analytical, numerical, or approximate) is independent of PhysGRIN. Therefore, in what follows, we present specific analytical results in the simplest possible form to validate the subsequent numerical integration of photon trajectories, as this approach does not require more elaborate mathematical notions or representations, which are beyond the scope of the present work.}

\begin{figure}[!ht]
  \centering
  \includegraphics[width=0.7\linewidth]{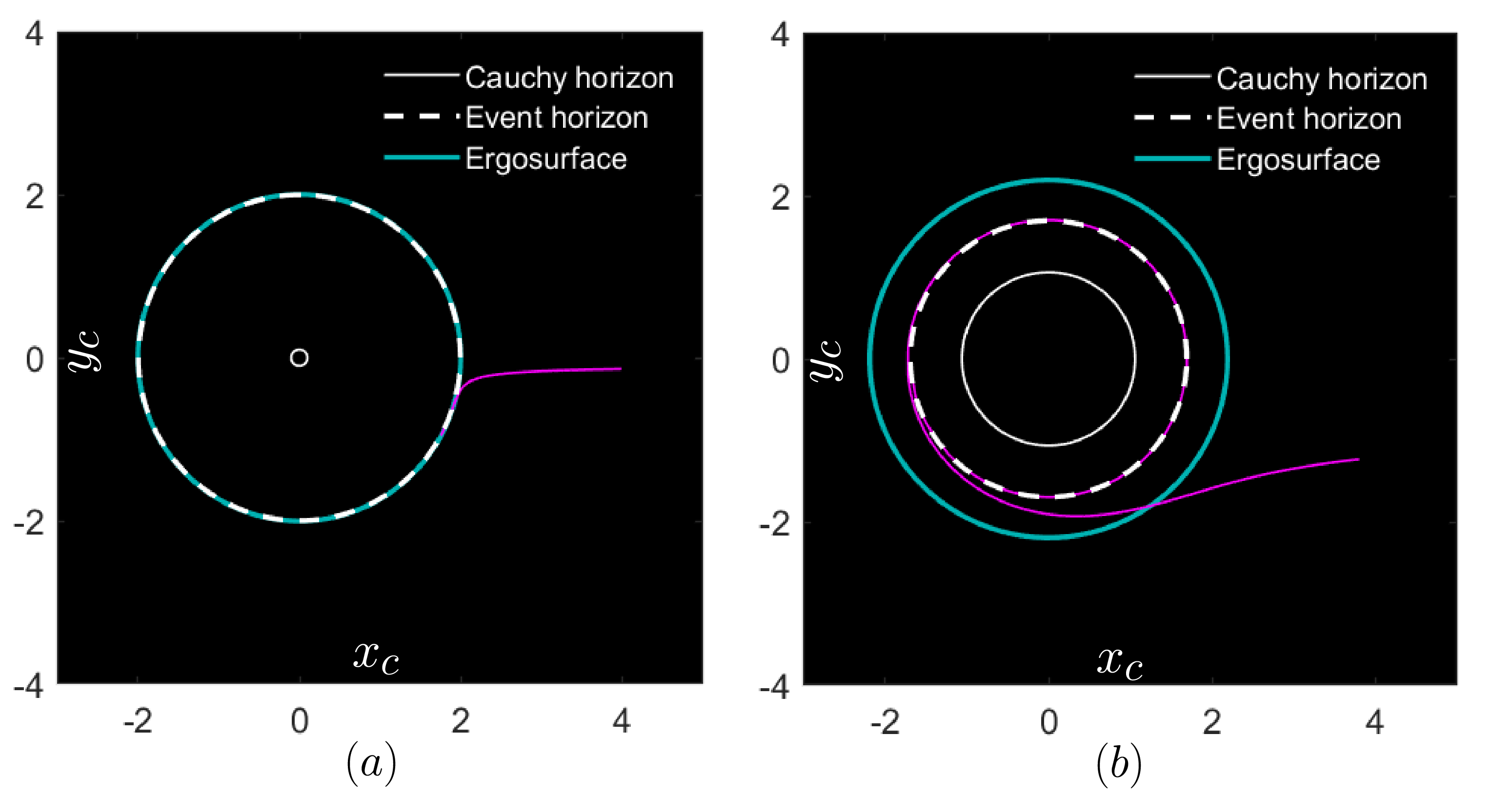}
  \caption{\textcolor{black}{ Geodesics with vanishing Carter constant are shown for the cases $(a)$ $a = 0.1$ and $(b)$ $a = 0.9$, using Planck units and $ m_b = 1$. Cartesian coordinates $x_c$ and $y_c$ were used.}}
  \label{fig:1}
\end{figure}

For convenience, let us define $ m_b \equiv GM / c^2 $ and $ T = ct $. The geodesics lying in the equatorial plane must satisfy \textcolor{black}{$ \theta = \pi /2 $, $ \dot{\theta} = 0 $, and $ \ddot{\theta}=0$}. A direct insight is that the Kerr metric (and thus the Lagrangian) does not depend explicitly on $ t $ and $ \phi $. Therefore, we have the following two integrals \cite{Chandrasekhar1983mathematical}:
\begin{eqnarray}
\nonumber
    \left( 1 - \frac{2m_b}{r_{bl}} \right) \dot{T} + \frac{2m_b a}{r_{bl}} \dot{\phi} = \frac{E}{c} \equiv \varepsilon,\\ \label{eq_sys}
    \left( r_{bl}^2 + a^2 + \frac{2m_b a^2}{r_{bl}} \right) \dot{\phi} - \frac{2m_b a}{r_{bl}} \dot{T} \equiv L,
\end{eqnarray}
where \( E \) and \( L \) denote the energy per unit mass and the angular momentum per unit mass, respectively. {\color{black}Thus, from Eqs. (\ref{eq_sys}), \( \dot{T} \) and \( \dot{\phi} \) can be expressed in terms of \( r_{bl} \), \( L \), and \( \varepsilon \). Substituting these relations into the null condition \( \textcolor{black}{\mathrm{d}}s^2 = 0 \) (equivalently, \( \mathcal{L} = 0 \)), one obtains the following equation for \( \dot{r}_{bl} \) \cite{Chandrasekhar1983mathematical}:}
\begin{eqnarray}
    r_{bl}^{2} \dot{r}_{bl}^2 = (r_{bl}^2 + a^2) \varepsilon^2 - L^2 + \frac{2m_b}{r_{bl}} \left( a \varepsilon - L \right)^2.
\end{eqnarray}
Since we are only interested in the trajectories of photons, we can eliminate the dependence on the affine parameter as $ \dot{\phi}/\dot{r}_{bl} = \textcolor{black}{\mathrm{d}}\phi/ \textcolor{black}{\mathrm{d}}r_{bl} $, which can be rewritten as \cite{fan2018analytical} 
\begin{eqnarray}
\label{dr}
    \textcolor{black}{\mathrm{d}}\phi = \frac{hr_{bl}^2 - 2m_b r_{bl}(h - a)}{\Delta \sqrt{N}} \, \textcolor{black}{\mathrm{d}}r_{bl},
\end{eqnarray}
where
\begin{eqnarray}
    h=\frac{L}{\varepsilon},\quad N=r_{bl}^4 + r_{bl}^2(a^2 - h^2) + 2m_b r_{bl}(a - h)^2.
\end{eqnarray}
When $h=a$, the integration of Eq. (\ref{dr}) is straightforward \cite{Chandrasekhar1983mathematical}
\begin{eqnarray}
\label{radial_geodesics}
    \phi = \frac{a}{r_{bl,-} - r_{bl,+}} \ln\left( \frac{r_{bl} - r_{bl,-}}{r_{bl} - r_{bl,+}} \right).
\end{eqnarray}
\textcolor{black}{The resulting trajectories, characterized by the vanishing of the Carter constant $K$ \cite{Carter_1968} (i.e., $K = \varepsilon^2(a - h)^2$ at $\theta = \pi/2$ and $\dot{\theta}=0$, which represents a conserved quantity that arises from the unexpected separability of the Hamilton-Jacobi equation), fulfill a similar role to that of radial geodesics in Schwarzschild geometry \cite{Chandrasekhar1983mathematical}.}
Fig.~\ref{fig:1} shows \textcolor{black}{such} geodesics for a Kerr black hole rotating clockwise for $ a =0.1 m_b $, and $ 0.9 m_b $.

\begin{figure}[!ht]
  \centering
  \includegraphics[width=0.7\linewidth]{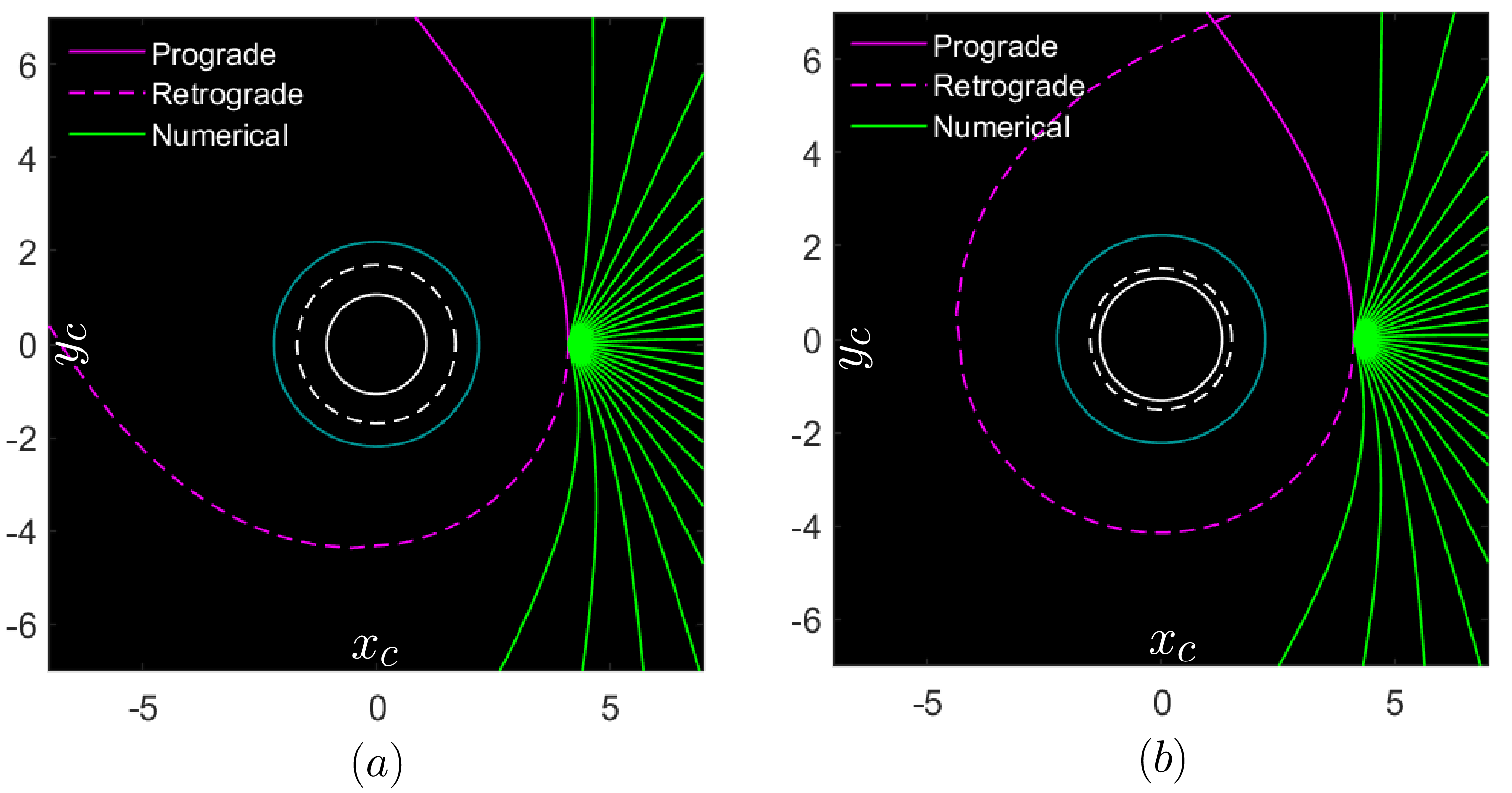}
  \caption{Exact prograde and retrograde paths and \textcolor{black}{numerically calculated} null geodesics for the cases $(a)$ $a=0.9$ and $(b)$ $0.99$, using Planck units and $m_b = 1$, and the point of convergence at $ r_{bl,0}=4 $. Cartesian coordinates $x_c$ and $y_c$ were used.}
  \label{fig:2}
\end{figure}

{\color{black}Equation (\ref{dr}) can be solved exactly using elliptic integrals, as first shown in the context of geodesic motion in the Schwarzschild metric by Hagihara \cite{hagihara1930theory} and later discussed in Chandrasekhar's classical treatment \cite{Chandrasekhar1983mathematical}. Specifically, Equation (\ref{dr}) can be integrated exactly in terms of elliptic integrals of the first and third kinds \cite{fan2018analytical,gradshteyn2000table}} when integrating from the point $r_{bl,0}$, where the distance from the geodesic to the event horizon is minimized,  to the position $r_{bl}$, with $r_{bl} > r_{bl,0}$. By reparametrizing $r_{bl}=2m_{b}x$, $a=2m_{b}j$, and $h=2m_{b}\psi$, where $x, j, \psi$ are dimensionless, this condition of minima distance is equivalent to \cite{fan2018analytical}  
\begin{eqnarray}
    f(x_0)=x_0^4 + (j^2 - \psi^2)x_0^2 + (j - \psi)^2 x_0 = 0.
\end{eqnarray}
Thus, for a given value of $x_0$,  $\psi$ is determined \cite{fan2018analytical}:  
\begin{eqnarray}
\label{psi}
    \psi_{\pm} = \frac{-j \pm x_0 \sqrt{x_0^2 - x_0 + j^2}}{x_0 - 1},
\end{eqnarray}
where $\pm$ indicates prograde (+) and retrograde (-) trajectories. The explicit form of the exact solution to these null geodesics can be found in Ref.~\cite{fan2018analytical}. However, it is important to note a misprint in Eq. (17) in that article: the variable $z$ should be replaced with $\sin(z)$. Under these conditions, it is straightforward to demonstrate that when $j = 0$, the results in Ref.~\cite{fan2018analytical} reduce to the Schwarzschild case, as shown in Ref.~\cite{Bret_n_2017}. Fig.~\ref{fig:2} presents the analytical prograde and retrograde geodesics for the cases $(a)$ $a = 0.1m_b$ and $(b)$ $a = 0.9m_b$ with $r_{bl,0} = 4m_b$.

 As a first attempt to extend our analysis, Eq. (\ref{dr}) is integrated numerically using the trapezoidal method, selecting values of $\psi$ within the range $(\psi_{-}, \psi_{+})$ that ensure the coefficient in Eq. (\ref{dr}) remains real-valued throughout the interval $(x_0, x)$. Fig.~\ref{fig:2} shows the numerically computed paths for the cases $(a)$ $a = 0.1m_b$ and $(b)$ $a = 0.9m_b$, with $r_{bl,0} = 4m_b$. It follows that these numerical paths escape from the black hole.

\begin{figure}[!ht]
  \centering
  \includegraphics[width=0.7\linewidth]{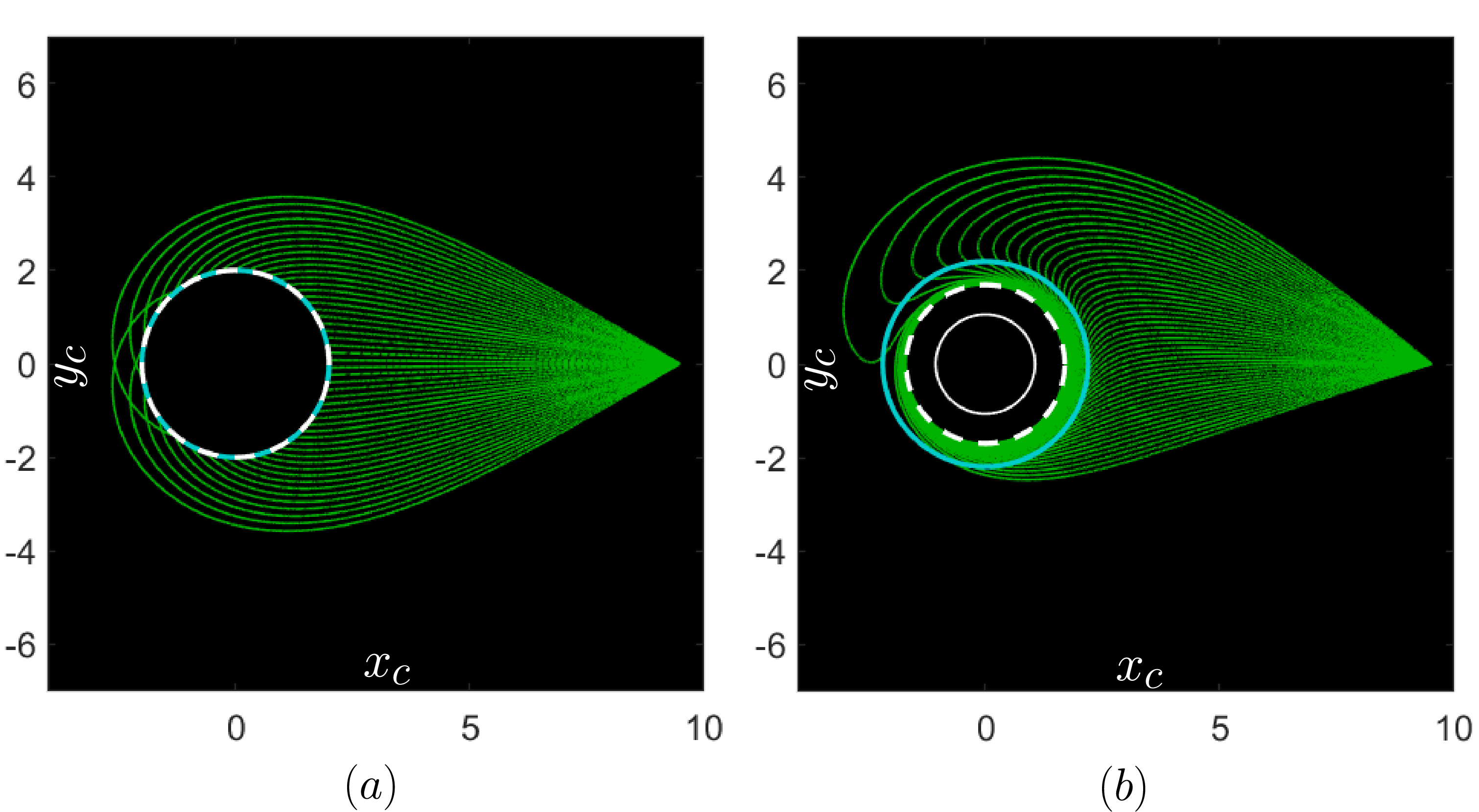}
  \caption{ \textcolor{black}{Numerically calculated} null geodesics for the cases $(a)$ $a=0$  and $(b)$ $0.9$, using Planck units and $m_b=1$, and the position of convergence at $r_{bl,f}=9.5$.  Cartesian coordinates $x_c$ and $y_c$ were used.}
  \label{fig:3}
\end{figure}

{\color{black}As we are interested in reproducing light trajectories around a black hole using a symmetric GRIN medium, we will focus on null geodesics that approach the event horizon and experience strong deflection. In Boyer–Lindquist coordinates, light rays do not cross the event horizon in finite coordinate time; instead, they asymptotically approach it. This coordinate behavior does not affect the physical equivalence of the modeled trajectories in the GRIN medium.} To obtain such trajectories through numerical integration, we first need to ensure that the coefficient of $\textcolor{black}{\mathrm{d}}r$ in Eq. (\ref{dr}) is a real-valued function, which implies that the values of $\psi$ must be restricted. This restriction arises from the condition $f(x) > 0$ for the interval $x \in (0, \infty)$, which can be reduced to 
\begin{eqnarray}
\label{g_condition}
   g(x) \equiv x^3 + (j^2 - \psi^2)x + (j - \psi)^2 > 0.
\end{eqnarray}
Since $g(x)$ is a cubic function of $x$ with a positive coefficient for $x^3$, the inequality Eq. (\ref{g_condition}) is satisfied if the evaluation of $g(x)$ at the critical value
\begin{eqnarray}
    x_{critical} = \frac{\sqrt{\psi^2 - j^2}}{\sqrt{3}},
\end{eqnarray}
is positive. The roots of $g(x_{critical}) = 0$ lead to the condition
\begin{eqnarray}
    (j - \psi)^2 = \frac{2(\psi^2 - j^2)^{3/2}}{3\sqrt{3}}.
\end{eqnarray}
Thus, by solving for $\psi$, \textcolor{black}{it is found that the range of values ensuring the well-defined numerical integration of Eq. (\ref{dr}) is given by: }
\begin{eqnarray}
\label{interval}
    \psi \in (\psi_1(j), \psi_2(j)),
\end{eqnarray}
where
\begin{eqnarray}
    \psi_1(j) &=& -j - \frac{3(1 + i\sqrt{3})}{4B(j)} - \frac{3}{4}\left(1 - i\sqrt{3}\right)B(j),\\
    \psi_2(j) &=& \frac{1}{2}\left( -2j + \frac{3}{B(j)} + 3B(j)\right),
\end{eqnarray}
and $B(j)= (-2j + \sqrt{4j^2 - 1})^{1/3}$. Finally, to ensure that all geodesics converge at the point $r_{bl,f}$ for all possible values of $\psi$, we integrate as follows:
\begin{eqnarray}
\label{numerical_geodesic}
    \phi(x) = \int_{x_0}^{x} \frac{x'(\psi x' + j - \psi)}{(x'^2 + j^2 - x')\sqrt{f(x')}} \textcolor{black}{\mathrm{d}}x' - \phi_0,
\end{eqnarray}
where
\begin{eqnarray}
    \phi_0 = \int_{x_0}^{x_f} \frac{x'(\psi x' + j - \psi)}{(x'^2 + j^2 - x')\sqrt{f(x')}} \textcolor{black}{\mathrm{d}}x'.
\end{eqnarray}
Here, $x_0$ is chosen to be arbitrarily close to the exterior event horizon. In Fig. \ref{fig:3}, it is shown the numerically computed paths using the trapezoidal method for $a = 0m_b$ and $0.9m_b$ by taking different values of $\psi$ in the range $(\psi_1(j), \psi_2(j))$; all trajectories converge at $r_{bl_f}=9.5m_b$.

\section{Fermat's ray invariant for GRIN media with spherical symmetry}

{\color{black}Fermat's principle can be expressed using orthogonal generalized coordinates associated with an arbitrary curvilinear orthogonal coordinate system that depends on a parameter $\eta$ $(q_1(\eta),q_2(\eta),q_3(\eta))$ as follows:
\begin{eqnarray}
\nonumber
\delta \int_C \mathcal{L}(q_1,q_2,q_3;\dot{q_{1}},\dot{q_{2}},\dot{q_{3}},\eta)\text{d}\eta = 0, 
\end{eqnarray}
where, $\mathcal{L}(q_1,q_2,q_3;\dot{q_{1}},\dot{q_{2}},\dot{q_{3}},\eta)$ is the optical Lagrangian, $\dot{q}_i = \textcolor{black}{\textcolor{black}{\text{d}}} q_i /\textcolor{black}{\text{d}}\eta $, with $i=1,2,3$, and $C$ represents the light-ray trajectories parameterized by $\eta$.}
\begin{figure}[!ht]
\centering\includegraphics[width=0.7\linewidth]{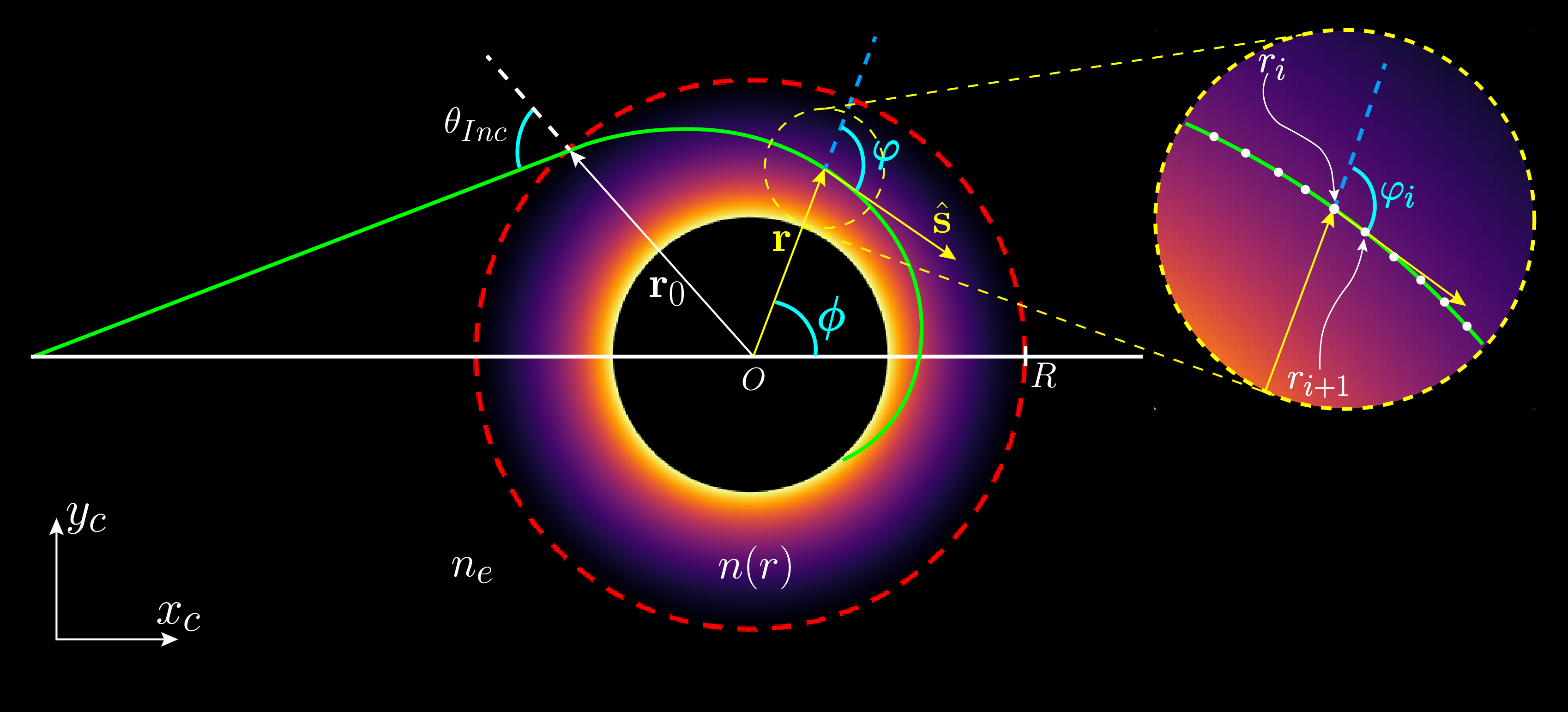}
\caption{The ray path propagating in a medium with a radially symmetric GRIN.}
\label{fig:4}
\end{figure}
When using a spherical coordinate system $(r, \theta, \phi)$ and a radially symmetric GRIN medium, it is sufficient to analyze the behavior of light along the equator $(\theta = \pi/2)$, since any trajectories lie in a plane passing through the origin. This is the primary reason why the equatorial null geodesics of a Kerr black hole will be modeled using a spherical GRIN medium thus, only a thin GRIN disk, where the light rays lie, is required \cite{Tinguely_2020}. Under these conditions, and by using $r$ as the parameter of the trajectories, the Lagrangian function is given by:
\begin{eqnarray}
\mathcal{L}(r) = n(r)\sqrt{1 + r^2 \phi_r^2},
\end{eqnarray}
with $\phi_r = \textcolor{black}{\text{d}}\phi /  \textcolor{black}{\text{d}}r $. Thus, the light-ray trajectories satisfy the Euler-Lagrange equations, and since the Lagrangian does not explicitly depend on $\phi$, there is a conserved quantity,  \textcolor{black}{$K_r\equiv \partial \mathcal{L}/\partial \phi_r$, known as Fermat's ray invariant, given explicitly by \cite{Gomez-Correa:21,Lakshminarayanan,G_mez_Correa_2023}:
\begin{equation}
K_r = r n(r) \sin{\varphi},
\label{ConstantKSnell}
\end{equation}
}
 where $\varphi$ is the angle between the vector $ \hat{{\bf s}}  $, tangent to the ray,  and $\bf{r}$, as illustrated in Fig. \ref{fig:4}. \textcolor{black}{In optics, Eq. (\ref{ConstantKSnell}) is referred to as generalized Snell's law for inhomogeneous media with spherical symmetry \cite{Luneburg_1964,Lakshminarayanan,Gomez-Correa:21, Gomez-Correa:22} or as Bouguer's formula \cite{Born_2019}. The value of $K_r$ is determined at the entrance point of the ray} \textcolor{black} {into the GRIN medium, at $r=R$.} {\color{black}It is important to emphasize that since $K_r$ is an invariant or conserved quantity, once defined it does not change as the light ray propagates.}

Eq. (\ref{ConstantKSnell}) shows that Fermat's ray invariant depends on variations in the GRIN distribution. Therefore, the GRIN distribution can be determined if the value of \textcolor{black}{$K_r$} is known and the trajectory is given (thus, $\varphi$ can be determined), as follows:

\begin{equation}
n(r) = \frac{K_r}{r \sin{\varphi}}.
\label{RefIndex}
\end{equation}

The expression above forms the core of the PhysGRIN method for spherical symmetry. The numerical implementation to determine the refractive index for photon trajectories around a Kerr black hole will be \textcolor{black}{discussed} in Sec. \ref{PhysGRIN_m}.

\section{Analytical GRIN reconstruction from photon trajectories}
\label{AnalyticalRec}

To determine the value of \textcolor{black}{$K_r$}, it is essential to know $\sin{\varphi}$. Consequently, as $\varphi$ is the angle between  $ \hat{\bf s}  $ and  $\bf{r}$, the following relationship holds:
\begin{eqnarray}
    \sin\varphi = \frac{r(\phi)}{\sqrt{r^2(\phi) + \left( \frac{\textcolor{black}{\text{d}}r}{\textcolor{black}{\text{d}}\phi} \right)^2}}.
    \label{ValuVarPhi}
\end{eqnarray}
By substituting Eq. (\ref{ValuVarPhi}) into Eq. (\ref{RefIndex}), the variation of the refractive index is given by:
\begin{eqnarray}
\label{exact_reconstruction}
    n(r) = \frac{K_r \sqrt{r^2 + (\textcolor{black}{\text{d}}r/\textcolor{black}{\text{d}}\phi)^2}}{r^2}.
\end{eqnarray}
 Now, let us assume that $n(R)=n_e$, and that $ \textcolor{black}{\text{d}}r/\textcolor{black}{\text{d}}\phi$ can be expressed as a function of $r$, that is, $\textcolor{black}{\text{d}}r/\textcolor{black}{\text{d}}\phi = q(r)$. From Eq. (\ref{exact_reconstruction}), at $r=R$, we can derive \textcolor{black}{$K_r$}:
\begin{eqnarray}
    K_r = \frac{R^2 n_e}{\sqrt{R^2+q^2(R)}}.
\end{eqnarray}
Then, Eq. (\ref{exact_reconstruction}) can be rewritten as:
\begin{eqnarray}
\label{n_q}
    n(r) = n_e \frac{R^2}{r^2} \sqrt{\frac{r^2+q^2(r)}{R^2+q^2(R)}}.
\end{eqnarray}
\textcolor{black}{Now, if we consider the Boyer-Lindquist coordinates in the equatorial plane ($\theta=\pi/2$), their relation to spherical coordinates is given by $r=\sqrt{r_{bl}^2 +a^2}$. Thus, we obtain:}

\begin{eqnarray}
\label{q_change}
   q(r) = \frac{r_{bl}}{\sqrt{r_{bl}^2+a^2}} \frac{\textcolor{black}{\text{d}}r_{bl}}{\textcolor{black}{\text{d}}\phi}.
\end{eqnarray}
\textcolor{black}{By substituting Eq. (\ref{dr}) into the last equation and Eq. (\ref{q_change}) into Eq. (\ref{n_q}), and then expressing the index of refraction in spherical coordinates, where $ r_{bl} = \sqrt{r^2 - a^2} $, we find that, in the most general case, the index of refraction is given by:
}

\begin{eqnarray}
\label{general_result}
   n(r) = n_e \frac{R^3}{r^3}\left|  \frac{e(R)}{e(r)} \right| \sqrt{ \frac{r^4 e^2(r) + s(r)}{R^4 e^2(R) + s(R)} },
   \label{Analytical_nr}
\end{eqnarray}
where
\begin{eqnarray}
s(r) &=& \left[r^2 -2m_b\sqrt{r^2-a^2}\right]^2  \left[ (r^2-a^2)(r^2-h^2)+2m_b\sqrt{r^2-a^2} (a-h)^2 \right],
\end{eqnarray}
and
\begin{eqnarray}
e(r) &=& h\sqrt{r^2-a^2} -2m_b(h-a).
\end{eqnarray}

 Equation (\ref{general_result}) establishes the analytical distribution of the refractive index when the trajectories are represented in Cartesian coordinates $x_c$, $y_c$. Thus, Eq. (\ref{general_result}) agrees with the observation made in Ref.~\cite{Ramezani_Aval_2024} that the most accurate definition of the refractive index is achieved when light rays in optical media coincide with null geodesics.

Our model considers only GRIN media reproducing null geodesic trajectories; it does not include other electromagnetic effects, such as polarization.

Although the same approach to obtaining the analytical form of the index of refraction was reported in Ref.~\cite{Tinguely_2020}, they do not provide an explicit form of the proportionality constant \textcolor{black}{$K_r$}. Moreover, the main difference of our procedure from theirs is that they represented the null geodesics in Boyer-Lindquist coordinates, thus avoiding the change of variables in Eq.~(\ref{q_change}), indicating that spherical coordinates and Boyer-Lindquist coordinates are not related by a conformal transformation (i.e., the GRIN medium depends on the coordinate system used to describe the trajectories). An alternative approach to obtaining the index of refraction that exactly reproduces photon trajectories for static spherically symmetric spacetimes involves introducing isotropic coordinates, where the curved spacetime becomes conformally flat \cite{Nouri_Zonoz_2022}. \textcolor{black}{However, this approach is not generally applicable for stationary spacetimes \cite{Parvizi_2024}; therefore, the approach presented here and the numerical method discussed in the next section offers a more general approach} to reconstruct GRIN media. Here, we decided to use Cartesian coordinates instead of Boyer-Lindquist coordinates because, in such a representation, the singularity of the Kerr black hole appears as an annulus, which is a common depiction.

From Eq. (\ref{n_q}), it is clear that the radial distribution of the GRIN medium depends on the nature of the function $ q(r) $. \textcolor{black}{The function's existence and explicit form} depend on the origin of the ray's trajectory. Determining this function can be particularly challenging if we focus on the paths of photons \textcolor{black}{lying in the equatorial plane of a black hole}. This task requires prior knowledge of the theory of relativity and tensor calculus, as well as an understanding of the nature and geometric properties of spacetime, as discussed in Sec. \ref{photon_trajectories}.

\section{Reconstruction without general relativity}
\label{PhysGRIN_m}

So far, in Sec. \ref{AnalyticalRec}, we have outlined the procedure to obtain the analytical form of the GRIN medium that reproduces null geodesics \textcolor{black}{ lying in the equatorial plane of a black hole}.
Obtaining $q(r)$ in a broader context requires understanding GR, which can be challenging for a non-expert. Moreover, the analysis can take significant time, which may not be necessary if the main goal is reconstructing the GRIN medium. We will now explain how using Eq. (\ref{RefIndex}) \textcolor{black}{enables a remarkable process simplification} when focusing the attention exclusively on the trajectory.

Recent findings demonstrate that the value of \textcolor{black}{$K_r$} can be determined independently of the GRIN distribution, as shown below \cite{Gomez-Correa:22}: 
\begin{equation}
K_r = R n_{e} \sin{\theta_{Inc}},
\label{KSurface2022}
\end{equation}
 where $n_{e}$ is the external refractive index in which the GRIN distribution is immersed, and $\theta_{Inc}$ is the angle of incidence of the straight light ray over the GRIN surface at the position ${\bf r}_0$ ($|{\bf r}_0|=R$), as shown in Fig. \ref{fig:4}.

Therefore, if the trajectory of a light ray is known, from Eqs. (\ref{ConstantKSnell}) and (\ref{KSurface2022}), it follows that the refractive index of the medium that reproduces such a trajectory can be calculated using the following equation:
\begin{equation}
n(r) = \frac{R n_{e} \sin{\theta_{Inc}}}{r \sin{\varphi(r)}}.
\label{Recons}
\end{equation}

The trajectory from which we will take inspiration to obtain the GRIN medium can come from previous research on null geodesics. These geodesics may be described using different parameterizations from those presented in this paper or may involve more complex mathematical or algorithmic methods to compute them. In earlier works, null geodesics could also result from specialized treatments of orbital families. In all these cases, we do not need to understand the detailed calculations behind these trajectories to reconstruct the GRIN medium that matches photon trajectories. We only need the ray's path. In other contexts, such as the motion of mechanical particles, trajectories may be based on experimental data.

We assume the ray is discretized to develop a systematic method for reconstructing the GRIN. Consider a discretized ray passing through a spherical GRIN lens, as illustrated in Fig. \ref{fig:4}. Let $ {\bf r}_{i} = (x_{i}, y_{i}) $ represent the points along the ray path, where $ i $ denotes the discretization index. According to Eq. (\ref{Recons}), the value of the index of refraction \textcolor{black}{at the point} ${\bf r}_i$ along the ray's trajectory is given by
\begin{eqnarray}
\label{n_n}
n(r_i) = \frac{R n_{e} \sin{\theta_{Inc}}}{r_i \sin{\varphi(r_i)}},
\label{n_ri}
\end{eqnarray}
where $r_i = \sqrt{x_i^2 + y_i^2}$. This expression can be quickly evaluated when the trajectory is known because $\varphi(r_i)$ \textcolor{black}{can be calculated from} \cite{G_mez_Correa_2023}
\begin{eqnarray}
\label{varphi_i}
\varphi_i = \arccos \left( \hat{{\bf r}}_i \cdot \hat{{\bf s}}_i \right),
\end{eqnarray}
where
\begin{eqnarray}
 \hat{{\bf r}}_i = \frac{(x_i, y_i)}{r_i}, \quad \hat{{\bf s}}_i = \frac{(x_{i+1}-x_i, y_{i+1}-y_i)}{\sqrt{(x_{i+1}-x_i)^2 + (y_{i+1}-y_i)^2}}.
\end{eqnarray}

Evaluating Eq. (\ref{n_ri}) at each point $ r_{i} $ enables the construction of an exact numerical method, referred to as the Physical GRIN Reconstruction (PhysGRIN) method \cite{G_mez_Correa_2023}. The accuracy of the results depends on the proximity between $ r_{i} $ and $ r_{i+1} $, as well as the precision of the computational tools used. The robustness of this approach allows the GRIN distribution to be reconstructed from a ray propagating around Kerr black holes. Since the physical system is not approximated, the PhysGRIN method is considered exact. This method is likely the most straightforward approach to reconstructing GRIN media in spherically symmetric cases. Furthermore, it does not require an in-depth understanding of the theory behind null geodesics; instead, implementing the algorithm is sufficient.

In Sec. \ref{Results}, the PhysGRIN method is applied to reconstruct the GRIN medium representing \textcolor{black}{an equatorial slice of the spacetime} of a Kerr black hole. These reconstructions are then compared to the analytical solution provided by Eq. (\ref{Analytical_nr}).

\section{Results}
\label{Results}

The photon trajectories in Figs. \ref{fig:5}–\ref{fig:7} $(a)$ and $(d)$ were calculated numerically using Eq. (\ref{numerical_geodesic}) in Planck units with $m_b = 1$, employing the trapezoidal method with 10,000,000 regular divisions and choosing $x_0 = 0.001 + 0.5 r_{bl,+}$. In all cases, the geodesic ends at $R = 4$.

The index of refraction in Figs.~\ref{fig:5}–\ref{fig:7} $(b)$, $(c)$, $(e)$, and $(f)$ was evaluated by considering the GRIN medium immersed in vacuum $(n_e = 1$). Although the analytical expressions of the index of refraction in Eq. (\ref{general_result}) can be evaluated within the exterior of the event horizon, only the trajectories outside this region are necessary when addressing gravitational lensing effects. Thus, we avoid the interior of the black hole.

 In Fig. \ref{fig:5}, we present $(a)$ a prograde trajectory $(h = 2.8444$) and $(d)$ a retrograde trajectory $(h = -6.832$) for a Kerr black hole with angular momentum $a = 0.9$. Based on these trajectories, the analytical refractive index distributions are given by Eq. (\ref{general_result}).

The refractive index distribution for the prograde trajectory is shown in Fig. \ref{fig:5}$(b)$, while that for the retrograde trajectory is depicted in Fig. \ref{fig:5}$(e)$. Figs. \ref{fig:5}$(c)$ and $(f)$ compare the analytical distributions with the numerical reconstructions obtained using the PhysGRIN method for both trajectories.

To evaluate the overall accuracy of the numerical reconstruction, we use the normalized root mean square error (NRMSE), defined as: \begin{equation} \text{NRMSE} = \sqrt{\frac{\sum_{i=1}^{m}\arrowvert{n\left(r_{i}\right)}-{\hat{n}\left(r_{i}\right)}\arrowvert^{2}}{\sum_{i=1}^{m}\arrowvert{n\left(r_{i}\right)}\arrowvert^{2}}}, \label{NRMSError} \end{equation} where $n(r_{i})$ represents the analytical distribution \textcolor{black}{at the point} $r_{i}$, $\hat{n}(r_{i})$ is the corresponding numerical reconstruction and $m$ is the total number of discretization points ($i=1, 2, ..., m$). Lower NRMSE values indicate less residual variance, meaning a better approximation. Here, we choose $m = 10,000$.

 The NRMSE is $9.4615 \times 10^{-5}$ for the prograde trajectory. For the retrograde trajectory, the index of refraction was limited to $n < 4$ because it diverges at the position where the trajectory transitions from retrograde to prograde, causing significant numerical errors; in such conditions, the NRMSE is $2.6884 \times 10^{-4}$.

\begin{figure}[!ht]
  \centering
  \includegraphics[width=0.5\linewidth]{Figure_5.png}
  \caption{\textbf{Kerr case:} Null geodesics for $a=0.9$ with $(a)$ $h=2.844$  and $(d)$ $h=-6.832$, using Planck units and $m_b=1$. The corresponding GRIN media are shown in $(b)$ and $(e)$. Figures $(c)$ and $(f)$ compare the exact gradient index (blue line) and PhysGRIN (green dashed line). The GRIN medium is reconstructed between the outer event horizon and $R=4$. The white line indicates the Cauchy horizon. }
  \label{fig:5}
\end{figure}

Another interesting case arises when $a = 0$ {\color{black}in Eq. (\ref{general_result}), which corresponds} to the Schwarzschild scenario. In this case, the refractive index distribution is given by:
\begin{eqnarray}
\label{grin_sch}
    n_s(r) = n_e \sqrt{\frac{1+2mh^2/r^3}{1+2mh^2/R^3}}.
\end{eqnarray}
This result agrees with Ref.~\cite{Tinguely_2020}, although they do not give the explicit form of the constant coefficient. 

 In Fig. \ref{fig:6}, we present the trajectories for $(a)$ $h = 1$ and $(d)$ $h = 5.196$. The refractive index distributions for these trajectories are shown in Fig. \ref{fig:6} $(b)$ and $(e)$, respectively. Figs. \ref{fig:6} $(c)$ and $(f)$ compare the analytical distributions with the numerical reconstructions obtained using the PhysGRIN method for both trajectories. For $h = 1$, the NRMSE is $5.3385 \times 10^{-6}$, and for $h = 5.196$, it is $2.8792 \times 10^{-5}$.

\begin{figure}[!ht]
  \centering
  \includegraphics[width=0.5\linewidth]{Figure_6.png}
  \caption{\textbf{The Schwarzschild case:} Null geodesics for $(a)$ $h=1$  and $(d)$ $h=5.196$, using Planck units and $m_b=1$. The corresponding GRIN media are shown in $(b)$ and $(e)$. Figures $(c)$ and $(f)$ compare the exact gradient index (blue line) and PhysGRIN (green dashed line). The GRIN medium is reconstructed between the Schwarzschild radius $r_c =2$ and $R=4$. 
 }
  \label{fig:6}
\end{figure}

Finally, a particular case to be considered, \textcolor{black}{since its analytical expression can be obtained straightforwardly, is when the geodesics have a vanishing Carter constant } ($a=h$) described by Eq. (\ref{radial_geodesics}). These are associated with the following index of refraction:
\begin{eqnarray}
\label{grin_radial}
    n_r(r) = n_e \frac{R^3}{r^3}\sqrt{\frac{r^4a^2 + \left(r^2 -2m_b\sqrt{r^2-a^2}\right)^2(r^2-a^2)}{R^4a^2 + \left(R^2 -2m_b\sqrt{R^2-a^2}\right)^2(R^2-a^2)}}.
\end{eqnarray}

 In Fig. \ref{fig:7}, the trajectories for $(a)$ $a = 0.1 $ and $(d)$ $a = 0.9 $ are presented. A comparison of Figs. \ref{fig:1} and \ref{fig:7} demonstrates that the numerical solution of Eq. (\ref{numerical_geodesic}) provides an adequate level of accuracy.

 The refractive index distributions for these trajectories are shown in Figs. \ref{fig:7} $(b)$ and $(e)$, respectively. Figs. \ref{fig:7} $(c)$ and $(f)$ compare the analytical distributions with the numerical reconstructions obtained using the PhysGRIN method for both trajectories. For $a=0.1$, the NRMSE is $5.3528 \times 10^{-5}$, and for $a=0.9$ is $5.1586 \times 10^{-5}$.

\begin{figure}[!ht]
  \centering
  \includegraphics[width=0.5\linewidth]{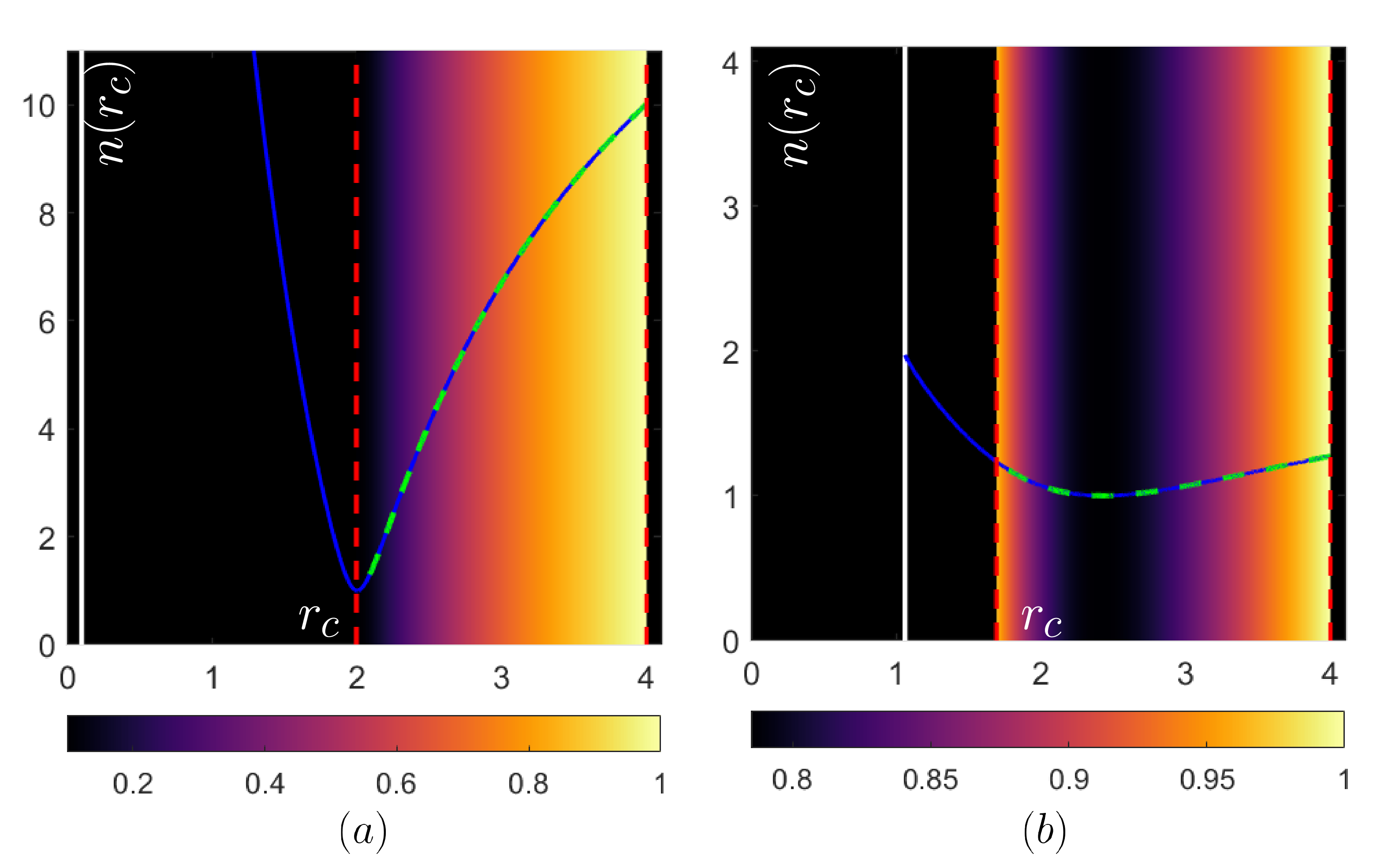}
  \caption{\textcolor{black}{The gradient index for geodesics with a vanishing Carter constant }($h=a$) for the cases $(a)$ $h=0.1$ and $n_e=1/0.0999$, and $(b)$ $h=0.9$ and $n_e=1/0.7850$, using Planck units and $m_b=1$. The GRIN medium is reconstructed between the outer event horizon and $R=4$. The white line indicates the Cauchy horizon.
 }
  \label{fig:8}
\end{figure}

Figs.~\ref{fig:7}~$(c)$ and $(f)$ show that, when the GRIN medium is considered to be in vacuum, the index of refraction must be less than 1, which would imply a speed of light greater than $c$. This issue can be addressed by immersing the GRIN in a homogeneous medium with an appropriate value of $n_e$. Figs.~\ref{fig:8} (a) and (b) illustrate the cases from Fig.~\ref{fig:7}~$(c)$ and $(f)$ but considering $n_e = 1/0.0999$ and $n_e = 1/0.7850$, respectively, thus ensuring that the minimum value of the index of refraction is equal to 1.

\begin{figure}[!ht]
  \centering
  \includegraphics[width=0.5\linewidth]{Figure_7.png}
  \caption{\textbf{Geodesics with a vanishing Carter constant ($h=a$):} Null geodesics for $(a)$ $h=0.1$ and $(d)$ $h=0.9$, using Planck units and $m_b=1$. The corresponding GRIN media are shown in $(b)$ and $(e)$. Figures $(c)$ and $(f)$ compare the exact gradient index (blue line) and PhysGRIN (green dashed line). The GRIN medium is reconstructed between the outer event horizon and $R=4$. The white line indicates the Cauchy horizon.
 }
  \label{fig:7}
\end{figure}

All the results presented in this section confirm that GRIN media effectively \textcolor{black}{reproduce} the null geodesic and demonstrate the feasibility of using a simple method for reconstructing the GRIN medium without relying on complex techniques. It only requires the ray trajectory \textcolor{black}{through the black hole spacetime}. This highlights the simplicity of the exact numerical method, which is one of the main objectives of the present work.

\section{Unstable GRIN media for radial null geodesics of the Schwarzschild black hole}

\textcolor{black}{In the Schwarzschild metric (\( a = 0 \)), a radial null geodesic is a straight line going to the center of the black hole $( \theta = \text{constant}$, $\phi = \text{constant} $). When the rays are considered to lie in the equatorial plane $( \theta = \pi/2 $) with the source at $ r = R $, $ \phi = 0$, numerical issues are expected when computing the GRIN medium for trajectories close to the radial geodesic (i.e., $ h \to 0 $), as in this case, $ K_r \to 0 $ in Eq. (\ref{KSurface2022}), which may introduce errors in the numerator of Eq. (\ref{n_n}) under such conditions.}

 On the other hand, from the analytical expression in Eq. (\ref{grin_sch}), when $h = 0$ and $ r \neq 0$, the index of refraction remains constant and equal to $n_e$, which is expected for a straight light ray moving toward the origin, interestingly, in Ref.~\cite{Tinguely_2020}, although their expression for the index of refraction is nearly identical, for the radial geodesic it tends to infinity throughout the space. This discrepancy arises because their unspecified constant of proportionality does not seem to depend on $h$.

On the other hand, when $h \neq 0$, Eq. (\ref{grin_sch}) predicts an infinite value for the index of refraction at $r = 0$, which is consistent with the fact that light cannot traverse or escape from the black hole. Mathematically, the index of refraction for the radial geodesic ($h = 0$) at $r = 0$ is undefined. One might be tempted to define the index of refraction for the radial geodesic by asymptotically approaching $h \to 0$, which would result in a constant index of refraction throughout space, except at $r = 0$ where its value becomes infinite. This appears to be consistent with a static black hole model. However, if we approach the radial geodesic by setting $a \to 0$ in Eq. (\ref{grin_radial}), we obtain:
\begin{eqnarray}
\label{correct_sch_rad}
    n_{s,r}(r) = n_e \frac{R^2}{r^2} \left| \frac{r^2 - 2m_b r}{R^2 - 2m_b R} \right|.
\end{eqnarray}
This expression has a completely different behavior than that derived from Eq. (\ref{grin_sch}) and yields an index of refraction of zero at the event horizon, which is not physically meaningful in optics. Furthermore, if we set $h = 0$ in Eq. (\ref{general_result}) and take the limit $a \rightarrow 0$, we obtain:
\begin{eqnarray}
    n(r) = n_e \frac{R}{r} \left| \frac{r^2 - 2m_b r}{R^2 - 2m_b R} \right|.
\end{eqnarray}
This results in yet another different expression for the index of refraction for the radial geodesic. In this case, the index of refraction is well-defined at $r = 0$, which is inconsistent with the black hole model as it allows light to traverse the black hole.

These discrepancies in the index of refraction when $h = 0$ and $a = 0$ demonstrate that there is no well-defined limit for the index of refraction for a radial geodesic in the \textcolor{black}{Schwarzschild spacetime}. \textcolor{black}{This indicates that the index of refraction is unstable} for radial geodesics. In experiments, this instability means that null geodesics close to the radial geodesics should be avoided.

\textcolor{black}{\section{Robustness of PhysGRIN under stochastic perturbations}}

\textcolor{black}{To evaluate the robustness of the PhysGRIN method, a null geodesic with vanishing Carter constant and parameters $h = a = 0.9$ is considered as a representative example. The geodesic starts at radius $x_0 = 0.001 + 0.5r_{bl,+}$, extends up to a radial distance of $R = 4$, and is discretized into $p$ data points. Each point is randomly displaced along the normal direction, introducing a standard deviation $\sigma$.}

\begin{figure}[!ht]
  \centering
  \includegraphics[width=0.5\linewidth]{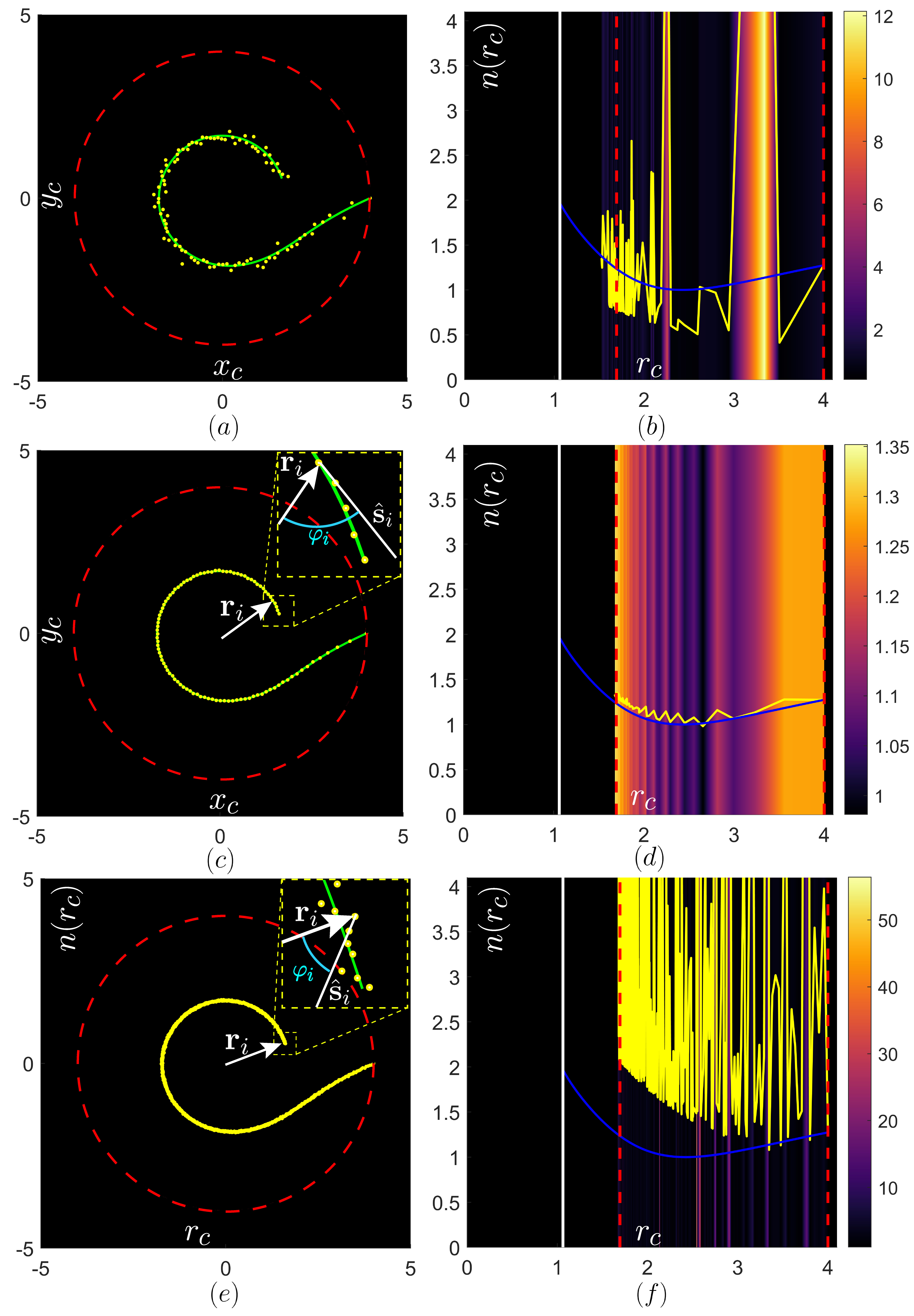}
  \caption{ \textcolor{black}{The exact null geodesic (green curve), with vanishing Carter constant and $a=h= 0.9$, is shown along with $p$ stochastic points (yellow) distributed with standard deviation $\sigma$ from the geodesic: $(a)$ $\sigma = 0.1$, $p = 100$; $(c)$ $\sigma = 0.01$, $p = 100$; and $(e)$ $\sigma = 0.01$, $p = 1000$. The corresponding reconstructed gradient index profiles are shown on the right (yellow curve and colormap) in figures $(b)$, $(d)$, and $(f)$, along with the exact gradient index (blue curve). Circular red dashed lines mark the exterior boundary of the GRIN media. Red dashed lines indicate the radial region where the exact null geodesic lies.  Planck units are used, and $m_b = 1$.}}
  \label{fig:9}
\end{figure}

\textcolor{black}{The PhysGRIN method relies on Eq.~(\ref{n_n}), which depends directly on the sine of the angle $\theta_{Inc}$ of the incident straight light ray over the GRIN surface with respect to its normal, and inversely on $\sin(\varphi_i)$, where $\varphi_i$ is defined in Eq.~(\ref{varphi_i}). Because of this, directly connecting the displaced points to form a trajectory and then applying the PhysGRIN method can lead to problems as shown in Fig~\ref{fig:9}. Figure~\ref{fig:9}$(a)$ shows the randomly displaced points (yellow dots) for $\sigma = 0.1$ and $p = 100$. The corresponding GRIN medium in Fig.~\ref{fig:9}$(b)$ yields an NRMSE of 1.1175 relative to the unperturbed case, indicating low reconstruction accuracy. Reducing the noise to $\sigma = 0.01$ while keeping $p = 100$ (Fig.~\ref{fig:9}$(c)$) significantly improves the result, with an NRMSE of 0.0667 (Fig.~\ref{fig:9}(d)). However, increasing the number of points to $p = 1000$ while keeping $\sigma = 0.01$ (Fig.~\ref{fig:9}$(e)$) leads to a significant drop in accuracy, with an NRMSE of 2.9081 (Fig.~\ref{fig:9}$(f)$).}

\textcolor{black}{When the angles $\varphi_i$ change smoothly between points, as seen in the zoomed view of Fig.~\ref{fig:9}(c), the index of refraction is estimated more accurately (Fig.~\ref{fig:9}(d)). However, abrupt changes in consecutive values of $\varphi_i$ cause large variations in the index of refraction. In extreme cases, when $\varphi_i$ approaches zero or $\pi$ (Fig.~\ref{fig:9}(e)), the index can become very large or even diverge (Fig.~\ref{fig:9}(f)). Moreover, $\theta_{Inc}$ is linked with the initial direction of the ray geodesic, and at this point, we know that for every trajectory we obtain a different index of refraction; thus, taking a random initial value of $\theta_{Inc}$ also leads to error.}

\begin{figure}[!ht]
  \centering
  \includegraphics[width=0.5\linewidth]{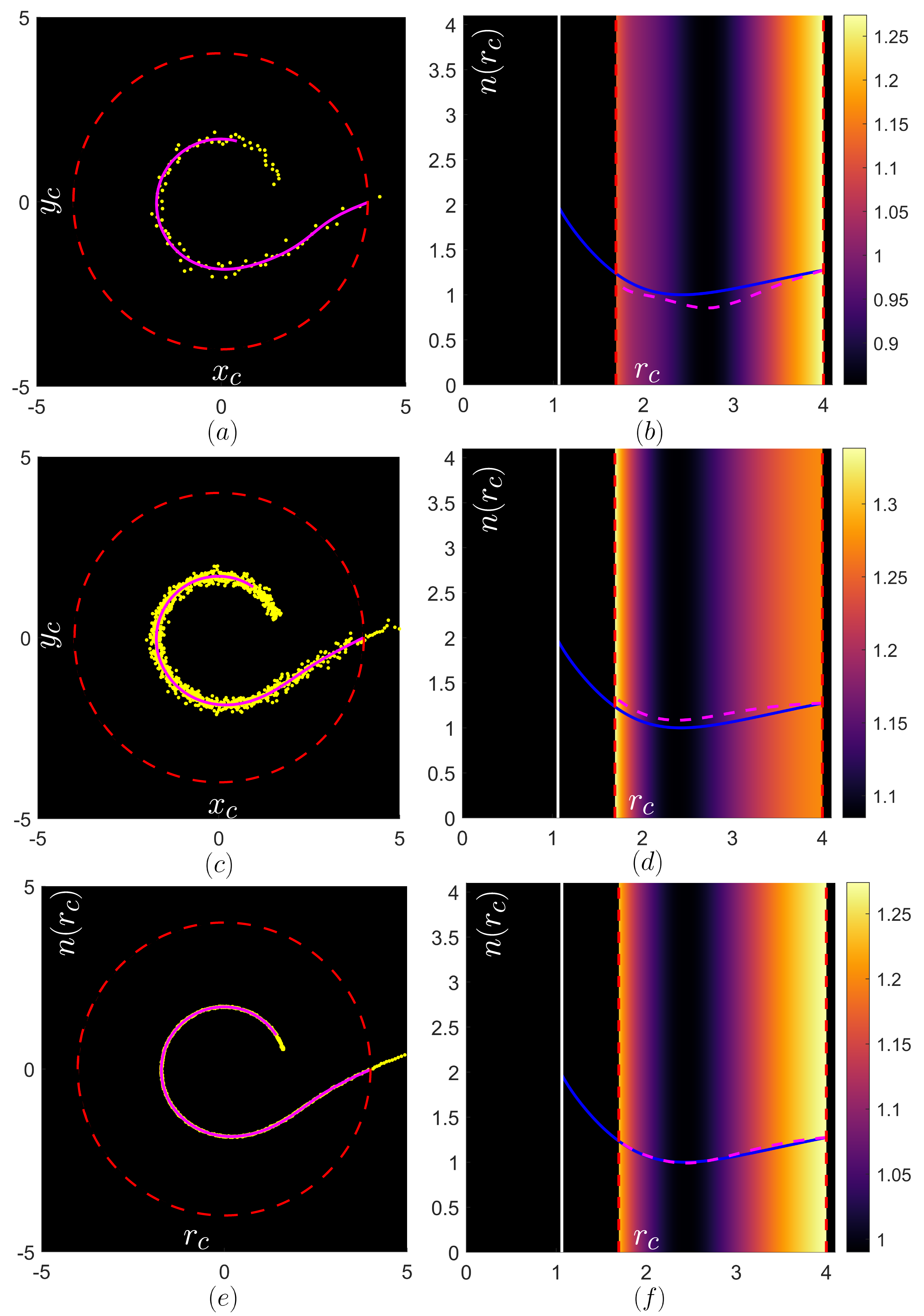}
  \caption{\textcolor{black}{The fitted geodesic (purple curve), obtained via least-squares minimization from $p$ stochastic points (yellow) with standard deviation $\sigma$ from the exact geodesic (with vanishing Carter constant and $a = 0.9$), is shown for: $(a)$ $\sigma = 0.1$, $p = 100$; $(c)$ $\sigma = 0.1$, $p = 1000$; and $(e)$ $\sigma = 0.01$, $p = 1000$. Figures $(b)$, $(d)$, and $(f)$ on the right show the corresponding reconstructed gradient index profiles (dashed purple curve and colormap), along with the exact gradient index profile (blue curve). Red dashed lines indicate the GRIN media boundaries. Planck units are used, and $m_b = 1$.} }
  \label{fig:10}
\end{figure}

\textcolor{black}{Taking control of this issue, or when dealing with stochastic data such as experimental trajectories of stars or planets (since PhysGRIN is directly applicable to massive particles, as will be shown in Section \ref{S2}), a suitable approach is to first fit a smooth curve to the data. This helps avoid abrupt changes in the index of refraction, which are not physically expected.}

\textcolor{black}{To build a smooth curve from the stochastic points, each data point $p$ is represented in polar coordinates as $(\theta_{\text{s},p}, r_{\text{s},p})$. The data suggest an exponential dependence on the angular coordinate $\theta_{\text{s},p}$, so the points are reparametrized as $(\exp(\theta_{\text{s},p}), r_{\text{s},p})$. The curve is then fitted using a fifth-degree polynomial via least-squares minimization \cite{burden2011numerical}.}

\textcolor{black}{The left column of Fig.~\ref{fig:10} shows the reconstructed ray (purple curve) from stochastic data for three cases: $\sigma = 0.1$ with $p = 100$ (Fig.~\ref{fig:10}$(a)$), $\sigma = 0.1$ with $p = 1000$ (Fig.~\ref{fig:10}$(c)$), and $\sigma = 0.01$ with $p = 1000$ (Fig.~\ref{fig:10}$(e)$). The corresponding GRIN media are shown on the right (Figs.~\ref{fig:10}$(b)$, $(d)$, and $(f)$). The NRMSE values are $0.0892$, $0.0844$, and $0.0056$, respectively. This clearly demonstrates a significant improvement, and therefore shows that the PhysGRIN method is robust against stochastic perturbations, provided that the trajectory is pre-smoothed (using, for instance, least-squares fitting), thereby enabling reliable and accurate reconstruction of complex GRIN media.}

\textcolor{black}{\section{Reconstructing the S2 Orbit via PhysGRIN: From Newtonian to Relativistic Models}}
\label{S2}
\textcolor{black}{
Up to this point, the capability of the PhysGRIN method to reconstruct the index of refraction profile that reproduces photon trajectories in the equatorial plane of a Kerr black hole has been demonstrated, including the particular \textcolor{black}{case of the Schwarzschild spacetime}. The method requires only the trajectory itself, making it independent of the technique used to obtain the path and, therefore, also independent of the nature of the particle; i.e., it can be applied to any object whose path through spacetime can be tracked.
}

\textcolor{black}{In this section, the capability and accuracy of the PhysGRIN method in distinguishing subtle relativistic deviations from Newtonian predictions in the motion of massive particles are evaluated through the analysis of the apsidal precession of the star S2, which orbits the nearest supermassive black hole candidate at the center of the Milky Way \cite{2020}. To distinguish this relativistic effect from the Newtonian gravitational potential case, three orbital models are considered: (i) a Newtonian elliptical orbit, (ii) a quasi-elliptical trajectory that includes precession, and (iii) the relativistic orbit obtained via numerical integration.}

\begin{figure}[!ht]
  \centering
  \includegraphics[width=0.7\linewidth]{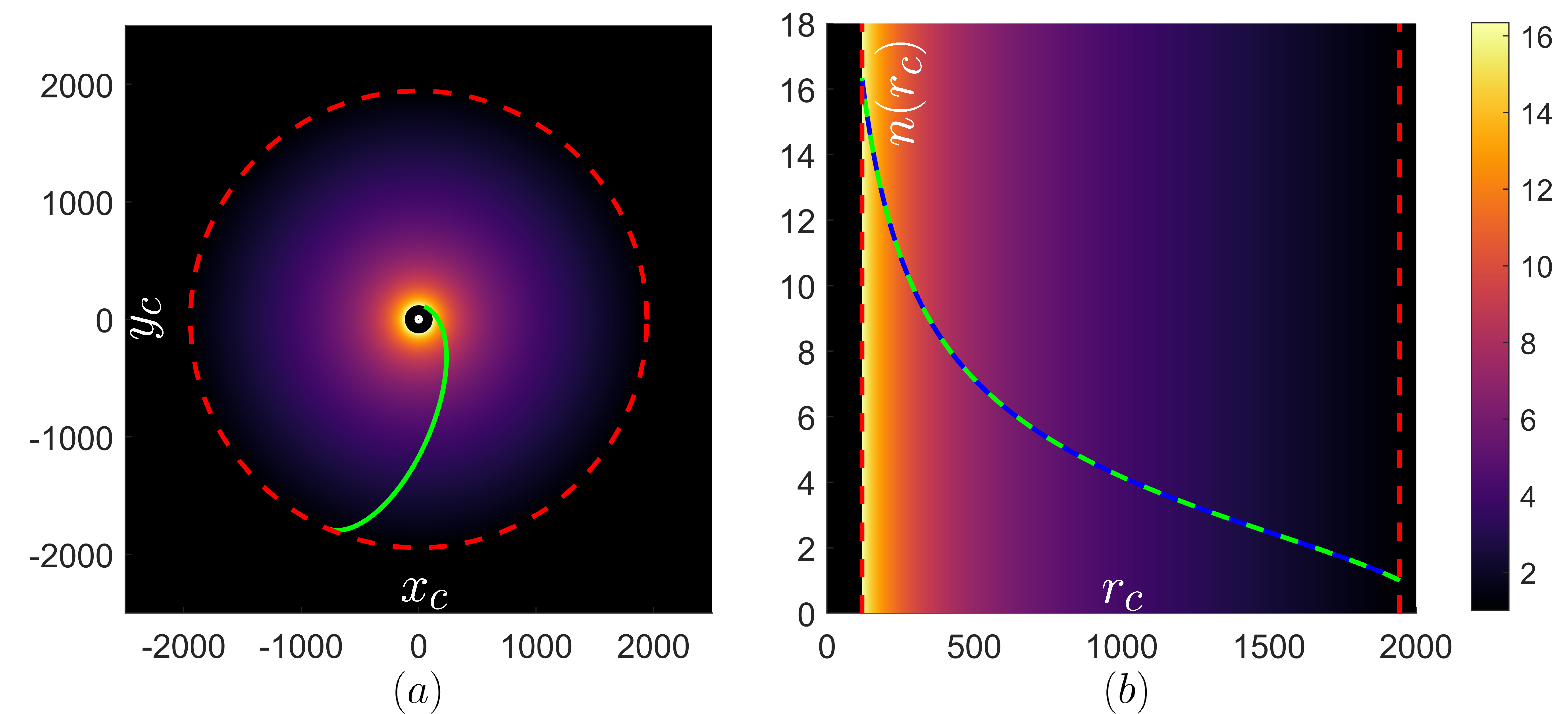}
  \caption{ \textcolor{black}{In figure $(a)$, half of the trajectory followed by the S2 star without precession is shown (green curve), along with the index of refraction profile obtained by applying the PhysGRIN method. Figure $(b)$ shows the exact index of refraction profile (blue curve) and the reconstructed profile (dashed green curve). Here, lengths are expressed in astronomical units.}}
  \label{fig:11}
\end{figure}

\textcolor{black}{ (i) {\bf The Newtonian elliptic orbit.} In short, within classical mechanics, the conserved energy per unit mass, $E_N$, and the orbital angular momentum per unit mass, $L_N$, for motion in the equatorial plane of a central potential satisfy \cite{goldstein2002classical}:}
\textcolor{black}{\begin{eqnarray}
\label{conserved_quantities}
    2E_N = \left( \frac{\textcolor{black}{\text{d}}r}{\textcolor{black}{\text{d}}t} \right)^2 + \frac{L_N^2}{r^2} + 2V(r), \quad L_N = r \frac{\textcolor{black}{\text{d}}\phi}{\textcolor{black}{\text{d}}t},
\end{eqnarray}
where, for an elliptical orbit in the Newtonian potential $V(r) = -GM/r$, the specific energy and angular momentum are \cite{goldstein2002classical}
\begin{eqnarray}
E_N = -\frac{GM}{2 a_{orb}} \quad \text{and} \quad L_N = \sqrt{GM a_{orb} (1 - e_{orb}^2)},
\end{eqnarray}}
\textcolor{black}{with $a_{orb}$ denoting the semi-major axis of the orbit and $e_{orb}$ its eccentricity. The position of the star S2 as a function of its angular position in the equatorial plane is given by the standard conic section equation\cite{goldstein2002classical}
\begin{eqnarray}
\label{eliptic_trajectory}
    r(\phi) = \frac{a_{orb}(1 - e_{orb}^2)}{1 + e_{orb}\cos\phi},
\end{eqnarray}}
 \textcolor{black}{Due to orbital symmetry, only half of the trajectory is sufficient to reconstruct the gradient index distribution using the PhysGRIN method, which reproduces the star’s path. By taking $M = 4.2997 \times 10^6,M_\odot$, $a = 1034.6$ AU, and $e = 0.88444$ \cite{Ashkenazy_2025}, the PhysGRIN method achieves a normalized root-mean-square error (NRMSE) of $7.3579 \times 10^{-7}$ with respect to the analytical index of refraction profile obtained through the optical-mechanical analogy \cite{leonhardt2010geometry, Xiao_2021}. This analytical profile is computed using the Newtonian potential and normalized such that the index of refraction equals $1$ at the apoapsis. Figure~\ref{fig:11} shows the reconstructed gradient index profile obtained with PhysGRIN, along with the exact index of refraction, revealing an almost perfect match. }

\textcolor{black}{(ii) {\bf Quasi-elliptical orbit accounting for precession}. As a first approach to modeling an orbit with apsidal precession, a representation of the trajectory is obtained through a quasi-elliptical path by replacing the term $\cos\phi$ in Eq.~(\ref{eliptic_trajectory}) with $\cos(\sqrt{K_p}\phi)$, following \cite{Hall_2022, rindler2006relativity}, where
\begin{eqnarray}
K_p = 1 - \frac{6GM}{c^2 a_{orb}(1 - e_{orb}^2)}.
\end{eqnarray}}
\textcolor{black}{Although this approximation is valid for orbits with low eccentricity, which deviates only slightly from a circular orbit. Such a modification yields an apsidal shift of
\begin{eqnarray}
2\pi\left(\frac{1}{\sqrt{K_{p}}} - 1\right) \approx 12.2'.
\end{eqnarray}}
\textcolor{black}{a value that does not differ significantly from the $12.1'$ reported in \cite{2020}. Such a precession effect is not easy to detect over a single orbital period, making it difficult to distinguish from an elliptical trajectory. In the context of an optical device with dimensions on the order of meters, reproducing precession would be particularly challenging. Although more precise values for the precession can be obtained by accounting for the stellar distribution around SgrA* \cite{Ashkenazy_2025}, the objective here is to assess whether sufficient accuracy can be achieved via the PhysGRIN method to differentiate between slightly different trajectories. Therefore, the present model is adequate for such analysis. }

\textcolor{black}{The PhysGRIN reconstruction for this quasi-elliptic orbit yields an NRMSE of $3.3219 \times 10^{-4}$ with respect to the exact index of refraction reported in \cite{Xiao_2021} for the modified Newtonian potential $V(r) = -GM r^{-1} - GM L^2c^{-2} r^{-3}$ \cite{rindler2006relativity}, thus exhibiting greater error than in the previous case of the Newtonian elliptic orbit (on the order of $10^{-7}$). This increased error can be attributed to the quasi-elliptic approximation used for modeling the orbit of S2.}

\textcolor{black}{ (iii) {\bf Relativistic orbit obtained via numerical integration.}} \textcolor{black}{From Eq.~(\ref{conserved_quantities}), the angular position as a function of radius is given by
\begin{equation}
    \phi(r) = \int_{r_I}^{r_f} \frac{L \, \textcolor{black}{\text{d}}r}{r^2 \sqrt{2E - L^2 r^{-2} - 2V(r)}}
\end{equation}
To account for the relativistic effect of apsidal precession, it is sufficient to use the modified Newtonian potential  in the previous integral. To avoid numerical issues when using the trapezoidal method for integration (due to complex values arising inside the square root) the integration limits are chosen as $r_I = 1949.6$ and $r_f = 119.5$, values close to the apoapsis and periapsis distances of the S2 star relative to SgrA*, respectively.
The NRMSE associated with this numerical trajectory, compared to the exact gradient index for the modified Newtonian potential \cite{Xiao_2021}, is $2.3133 \times 10^{-6}$, thus showing less error than the quasi-elliptic precession orbit.}

\begin{figure}[!ht]
  \centering
  \includegraphics[width=0.7\linewidth]{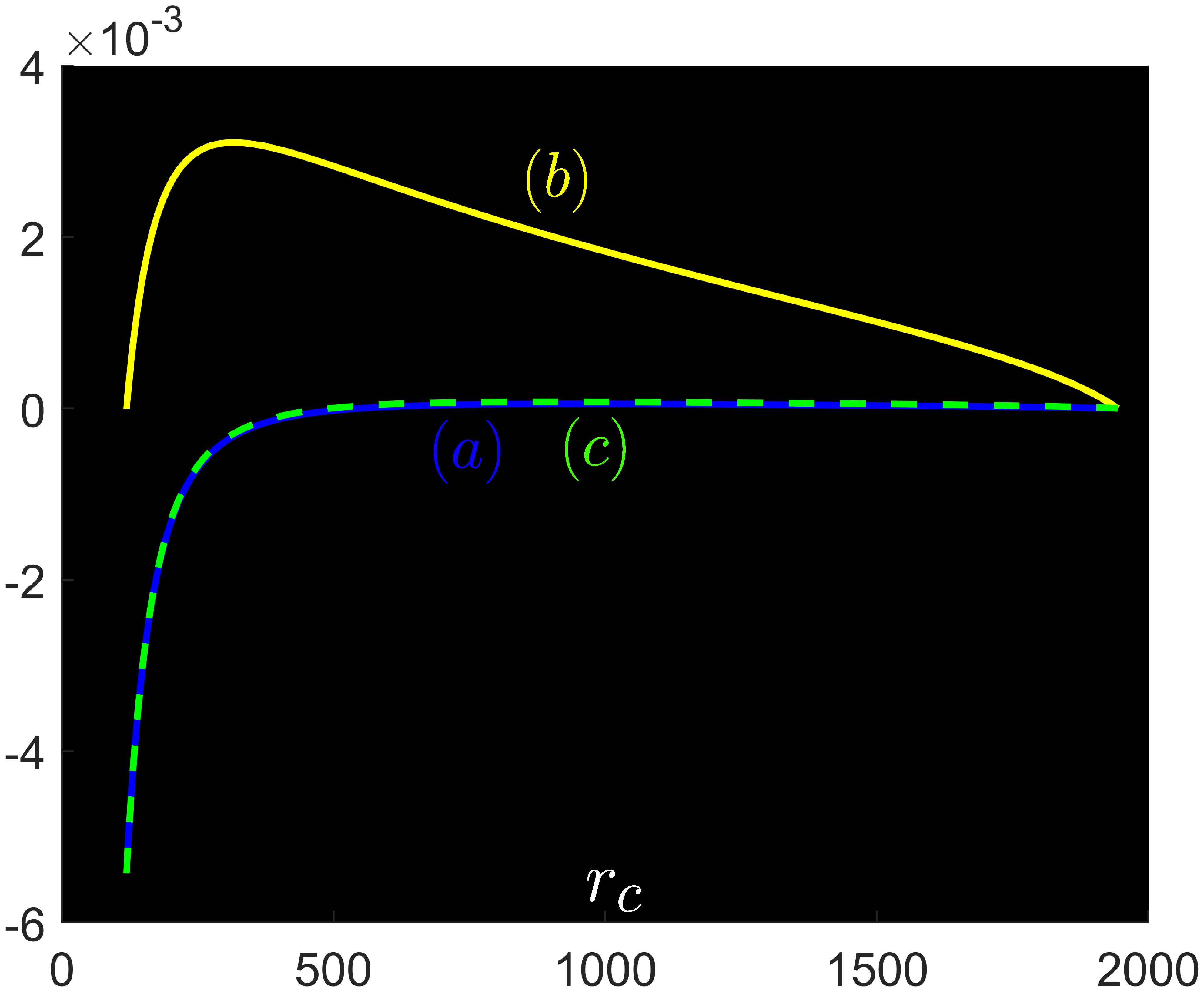}
  \caption{\textcolor{black}{Subtraction of refractive index profiles between the elliptical trajectory and those including apsidal precession for: (a) exact analytical indices from the optical-mechanical analogy~\cite{leonhardt2010geometry, Xiao_2021} (blue line); (b) PhysGRIN reconstructions of the elliptical and quasi-elliptical trajectories (yellow line); and (c) PhysGRIN reconstructions of the elliptical and numerically computed relativistic trajectories (green dashed line). In all cases, 1,000,000 points were used to generate the corresponding orbit trajectory. Here, length is expressed in astronomical units.}}
  \label{fig:12}
\end{figure}

\textcolor{black}{Figure~\ref{fig:12} shows the subtraction between the refractive index profiles corresponding to the elliptical trajectory and those including apsidal precession, for the following cases: (a) the subtraction of the exact analytical indices of refraction obtained via the optical-mechanical analogy~\cite{leonhardt2010geometry, Xiao_2021} \textcolor{black}{between} the Newtonian and modified Newtonian potentials (blue line); (b) the subtraction of the PhysGRIN reconstruction between the elliptical trajectory (case (i)) and the quasi-elliptical trajectory (case (ii)), shown by the yellow line; and (c) the subtraction of the PhysGRIN reconstructions between the elliptical trajectory (case (i)) and the numerically computed relativistic trajectory of the S2 star (case (iii)), shown by the green dashed line. Although all models yield low NRMSE values, the differences between curves (b) and (c) in Fig.~\ref{fig:12} are clearly noticeable. Resolving such variations requires reconstruction errors below $10^{-5}$. Therefore, in practice, fabricating a GRIN medium capable of accurately reproducing the apsidal precession of the S2 star would be challenging.
 }

\section{Conclusions}

In this work, we presented a methodology based on the numerical PhysGRIN method to derive GRIN media \textcolor{black}{replicating photon trajectories in the equatorial plane of the Kerr spacetime}. This approach leverages the system's symmetry and known photon trajectories, enabling the construction of GRIN media without requiring advanced knowledge of GR or tensor calculus \textcolor{black}{while remaining independent of the method used to calculate the null geodesics (whether analytic, numeric, or approximate)}.  

\textcolor{black}{While PhysGRIN relies on the conservation of Fermat's ray invariant via Noether's theorem (providing a robust and flexible numerical tool for modeling complex optical media), the theory presented here can be extended to model light rays outside the equatorial plane of the Kerr spacetime and potentially for more exotic metrics, using a three-dimensional freeform GRIN medium \cite{Lippman_2021}. Due to the complexity of this task, this topic will be addressed in a future paper.} \textcolor{black}{Note that although photon trajectories are not directly observable, gravitational lensing can give us indirect experimental verification of such theoretical trajectories \cite{Lemon_2024}; furthermore, we can also account for the deflection phenomena of massive particles due to the spacetime curvature using GRIN media, as shown in \cite{Xiao_2021}, or by applying the method presented in this work. Moreover, relativistic particles, such as neutrinos emitted by the accretion disk of a massive compact object, could help detect their shadows, as suggested in \cite{Mendoza_2024}. These trajectories can also be modeled using GRIN media with the approach presented in this paper.}

Additionally, by applying the conservation of Fermat's ray invariant, we derived an analytical form of the GRIN medium for equatorial null geodesics around Kerr black holes in Cartesian coordinates (we used Cartesian coordinates to represent ray trajectories as they represent the Kerr black hole singularity as an annulus, a widely accepted depiction). In contrast, Ref.~\cite{Tinguely_2020} provided expressions in Boyer-Lindquist coordinates, differing from ours except in the Schwarzschild case, where they coincide to a constant, leading to additional issues for radial geodesics. Moreover, the approach presented here goes beyond the limitation of mapping the spatial part of the metric to a conformally flat coordinate system to obtain the index of refraction that exactly reproduces photon trajectories, as suggested in \cite{Parvizi_2024}. Thus, PhysGRIN offers a more general approach to reconstruct GRIN media.

Our results align with the suggestion of Ref.~\cite{Ramezani_Aval_2024} that the most accurate definition of the index of refraction compares null geodesics in spacetime to light rays in optical media. This approach is effective when concentrating solely on ray trajectories without considering additional electromagnetic wave properties such as polarization.

\textcolor{black}{ An extension of this work would be the application of PhysGRIN to model light ray propagation around black holes surrounded by plasma. Plasma near a black hole alters the equations of motion for photons. One way to account for such deviations in photon trajectories is by introducing refractive \textcolor{black}{properties of the spacetime} through an index of refraction that depends on the position and \textcolor{black}{the frequency of the light} \cite{Atamurotov_2015, Rogers_2015}. If the plasma distribution confines light rays to the equatorial plane, the method introduced here to reconstruct GRIN media with spherical symmetry is applicable, but now in a non-normalized GRIN medium \cite{Gomez-Correa:22} (a normalized GRIN medium is understood when the refractive index at its surface matches that of the surrounding medium). On the other hand, when plasma is present, additional effects arise due to light-matter interactions. While we can conjecture that light-matter interactions could be modeled by incorporating dispersion into the GRIN medium,} \textcolor{black}{ it is important to note that light polarization (such as those arising from the presence of an accretion disk  \cite{Matt_1993,Beloborodov_1998}) is irrelevant in the context of light propagation in a two-dimensional plane, which is the case of light propagating in the equatorial plane of a Kerr black hole. However, when considering more general photon trajectories, a bianisotropic medium that reproduces null geodesics and accounts for polarization must be used \cite{Tinguely_2020}. Optical analogs of such astrophysical systems require further study, which is beyond the scope of this manuscript. }

{\color{black}Another phenomenon that could be modeled using GRIN media is the propagation of gravitational waves. In the high-frequency regime—when the wavelength is much smaller than the characteristic curvature scale of spacetime—gravitational waves propagate along null geodesics, analogous to light rays in geometrical optics. This approximation, developed by Isaacson in 1968 through a WKB-type expansion of Einstein’s field equations, remains valid even in strong gravitational fields and describes gravitational waves as spin-2 perturbations traveling on a curved background \cite{Isaacson1968}. Under these conditions, light treated as a scalar wave in a GRIN medium can effectively emulate their behavior.}

\textcolor{black}{Linking ray optics and wave optics through integral diffraction catastrophes is a common practice \cite{Berry_1980}; this, however,  must be treated with caution in the case of gravitational lensing for the following reasons: First, observing interference phenomena in real gravitational lensing scenarios is challenging due to coherence issues arising from the finite size of sources or the lack of near-perfect alignment, as mentioned in \cite{Rodr_guez_Fajardo_2023}. On the other hand, while gravitational waves might be better suited for observing diffraction and interference phenomena experimentally, they are only pertinent for gravitational waves with wavelengths comparable to the Schwarzschild radius; however, this is beyond geometric optics \cite{Cusin_2020}. Second, the GRIN media approach introduced in this paper depends on the trajectory of the photon; hence,  only narrow, highly directional beams (such as Gaussian beams) are expected to best approximate geodesic trajectories, as shown in  \cite{Tinguely_2020}. This means that wave phenomena can only be analyzed within a narrow region. Addressing the connection between ray optics and wave optics requires optical metamaterials or freeform inhomogeneous media capable of capturing the global behavior of photon trajectories.}

{\color{black}To conclude, the PhysGRIN method introduced in this work provides a general and systematic approach to constructing GRIN media that accurately reproduce null geodesics in curved spacetimes. By relying on the symmetries of light paths, the method enables the design of refractive index distributions without the need to map the spacetime metric to a conformally flat form, thus offering a versatile tool for modeling light propagation in both relativistic and analog optical systems. Moreover, its trajectory-based formulation allows the method to be extended to simulate particle dynamics in other physical contexts, including classical mechanics and quantum systems such as Bohmian trajectories. }\textcolor{black}{When dealing with stochastic data, the method remains robust by fitting a smooth curve that mitigates errors from spurious values in the expected refractive index caused by abrupt changes in ray direction. The approach is simple, practical, and sufficiently precise to distinguish subtle differences between trajectories, as shown in the comparison of orbital models with and without the apsidal precession of the S2 star around SgrA*.}


\begin{acknowledgments}
The corresponding author thanks the Secretaría de Ciencia, Humanidades, Tecnología e Innovación (SECIHTI) for the support provided through the awarded postdoctoral scholarship (CVU 640385).
\end{acknowledgments}


\bibliography{apssamp}   

\begin{thebibliography}{10}

\bibitem{Schneider_1992}
P.~Schneider, J.~Ehlers, and E.~E. Falco, {\em Gravitational Lenses}, Springer
  Berlin Heidelberg  (1992).

\bibitem{Einstein_1911}
A.~Einstein, ``Über den einfluß der schwerkraft auf die ausbreitung des
  lichtes,'' {\em Annalen der Physik} {\bf 340}, 898–908  (1911).

\bibitem{einstein1915perihelion}
A.~Einstein, ``Erkl{\"a}rung der perihelbewegung des merkur aus der allgemeinen
  relativit{\"a}tstheorie,'' {\em Sitzungsberichte der K{\"o}niglich
  Preu{\ss}ischen Akademie der Wissenschaften (Berlin)} , 831--839  (1915).
\newblock Presented on 18 November 1915.

\bibitem{Perlick_2004}
V.~Perlick, ``Gravitational lensing from a spacetime perspective,'' {\em Living
  Rev. Rel.} {\bf 7}  (2004).

\bibitem{Rodr_guez_Fajardo_2023}
V.~Rodríguez-Fajardo, T.~P. Nguyen, K.~S. Hocek, {\em et~al.}, ``Einstein
  beams and the diffractive aspect of gravitationally-lensed light,'' {\em New
  J. Phys.} {\bf 25}, 083033  (2023).

\bibitem{Bartelmann_2010}
M.~Bartelmann, ``Gravitational lensing,'' {\em Class. Quantum Grav.} {\bf 27},
  233001  (2010).

\bibitem{Grespan_2023}
M.~Grespan and M.~Biesiada, ``Strong gravitational lensing of gravitational
  waves: A review,'' {\em Universe} {\bf 9}, 200  (2023).

\bibitem{Raab_2017}
F.~J. Raab and D.~H. Reitze, ``The first direct detection of gravitational
  waves opens a vast new frontier in astronomy,'' {\em Curr. Sci} {\bf 113},
  657  (2017).

\bibitem{Castelvecchi_2019}
D.~Castelvecchi, ``Black hole pictured for first time — in spectacular
  detail,'' {\em Nature} {\bf 568}, 284–285  (2019).

\bibitem{Virbhadra_2000}
K.~S. Virbhadra and G.~F.~R. Ellis, ``Schwarzschild black hole lensing,'' {\em
  PRD} {\bf 62}  (2000).

\bibitem{Virbhadra_2009}
K.~S. Virbhadra, ``Relativistic images of schwarzschild black hole lensing,''
  {\em PRD} {\bf 79}  (2009).

\bibitem{Virbhadra_2022}
K.~Virbhadra, ``Distortions of images of {S}chwarzschild lensing,'' {\em PRD}
  {\bf 106}  (2022).

\bibitem{Virbhadra_2024}
K.~S. Virbhadra, ``Conservation of distortion of gravitationally lensed
  images,'' {\em PRD} {\bf 109}  (2024).

\bibitem{Virbhadra_2024_c}
K.~Virbhadra, ``Compactness of supermassive dark objects at galactic centers,''
  {\em Can J Phys} {\bf 102}, 523–528  (2024).

\bibitem{Johnson_2023}
M.~D. Johnson, K.~Akiyama, L.~Blackburn, {\em et~al.}, ``Key science goals for
  the next-generation event horizon telescope,'' {\em Galaxies} {\bf 11}, 61
  (2023).

\bibitem{Plebanski_1960}
J.~Plebanski, ``Electromagnetic waves in gravitational fields,'' {\em Phys.
  Rev.} {\bf 118}, 1396–1408  (1960).

\bibitem{de_Felice_1971}
F.~de~Felice, ``On the gravitational field acting as an optical medium,'' {\em
  Gen. Relativ. Gravit.} {\bf 2}, 347–357  (1971).

\bibitem{Sch_tzhold_2002}
R.~Schützhold and W.~G. Unruh, ``Gravity wave analogues of black holes,'' {\em
  Phys. Rev. D} {\bf 66}  (2002).

\bibitem{Philbin_2008}
T.~G. Philbin, C.~Kuklewicz, S.~Robertson, {\em et~al.}, ``Fiber-optical analog
  of the event horizon,'' {\em Science} {\bf 319}, 1367–1370  (2008).

\bibitem{Genov_2009}
D.~A. Genov, S.~Zhang, and X.~Zhang, ``Mimicking celestial mechanics in
  metamaterials,'' {\em Nat. Phys.} {\bf 5}, 687–692  (2009).

\bibitem{Chen_2010}
H.~Chen, R.-X. Miao, and M.~Li, ``Transformation optics that mimics the system
  outside a {Schwarzschild} black hole,'' {\em Opt. Express} {\bf 18}, 15183
  (2010).

\bibitem{leonhardt2010geometry}
U.~Leonhardt and T.~Philbin, {\em Geometry and light: the science of
  invisibility}, Courier Corporation, Mineola, New York  (2010).

\bibitem{Tinguely_2020}
R.~A. Tinguely and A.~P. Turner, ``Optical analogues to the equatorial
  {K}err–{N}ewman black hole,'' {\em Commun. Phys.} {\bf 3}  (2020).

\bibitem{_van_ara_2024}
P.~Švančara, P.~Smaniotto, L.~Solidoro, {\em et~al.}, ``Rotating curved
  spacetime signatures from a giant quantum vortex,'' {\em Nature} {\bf 628},
  66–70  (2024).

\bibitem{eddington1920report}
A.~S. Eddington, {\em Report on the relativity theory of gravitation}, Fleetway
  Press, Limited  (1920).

\bibitem{eddington1921space}
A.~S. Eddington, {\em Space, time and gravitation: An outline of the general
  relativity theory}, The University Press  (1921).

\bibitem{Gordon_1923}
W.~Gordon, ``Zur lichtfortpflanzung nach der relativitätstheorie,'' {\em Ann.
  Phys.} {\bf 377}, 421–456  (1923).

\bibitem{Kovner_1990}
I.~Kovner, ``Fermat principle in arbitrary gravitational fields,'' {\em
  Astrophys. J.} {\bf 351}, 114  (1990).

\bibitem{Parvizi_2024}
A.~Parvizi, H.~Forghani-Ramandy, E.~Rahmani, {\em et~al.}, ``Photon rings in
  the metamaterial analog of a gravitomagnetic monopole,'' {\em PRD} {\bf 110}
  (2024).

\bibitem{Ramezani_Aval_2024}
H.~Ramezani-Aval, ``A comparative study on the gravitational analog of the
  spacetime index of refraction,'' {\em Chin. J. Phys.} {\bf 88}, 69–76
  (2024).

\bibitem{kogan2003invariant}
I.~A. Kogan and P.~J. Olver, ``Invariant {E}uler--{L}agrange equations and the
  invariant variational bicomplex,'' {\em Acta Appl. Math.} {\bf 76}, 137--193
  (2003).

\bibitem{G_mez_Correa_2023}
J.~E. Gómez-Correa, A.~L. Padilla-Ortiz, J.~P. Trevino, {\em et~al.},
  ``Symmetric gradient-index media reconstruction,'' {\em Opt. Express} {\bf
  31}, 29196  (2023).

\bibitem{Price_2018}
R.~H. Price and K.~S. Thorne, ``Lagrangian vs {H}amiltonian: The best approach
  to relativistic orbits,'' {\em Am. J.Phys.} {\bf 86}, 678–682  (2018).

\bibitem{thorne2000gravitation}
K.~S. Thorne, C.~W. Misner, and J.~A. Wheeler, {\em Gravitation}, W. H. Freeman
  and Company, San Francisco  (2000).

\bibitem{landau1960mechanics}
L.~D. Landau and E.~M. Lifshitz, {\em Mechanics}, vol.~1, CUP Archive  (1960).

\bibitem{wambsganss1998gravitational}
J.~Wambsganss, ``Gravitational lensing in astronomy,'' {\em Living Rev. Rel.}
  {\bf 1}, 1--74  (1998).

\bibitem{2020}
R.~Abuter, A.~Amorim, M.~Bauböck, {\em et~al.}, ``Detection of the
  {S}chwarzschild precession in the orbit of the star {S}2 near the galactic
  centre massive black hole,'' {\em Astronomy \&amp; Astrophysics} {\bf 636},
  L5  (2020).

\bibitem{mcnamara1978instability}
J.~M. McNamara, ``Instability of black hole inner horizons,'' {\em Proc. Roy.
  Soc. Lon. A} {\bf 358}(1695), 499--517  (1978).

\bibitem{Dexter_2009}
J.~Dexter and E.~Agol, ``A fast new public code for computing photon orbits in
  a {K}err spacetime,'' {\em Astrophys. J} {\bf 696}, 1616–1629  (2009).

\bibitem{Yang_2013}
X.~Yang and J.~Wang, ``Ynogk: A new public code for calculating null geodesics
  in the {K}err spacetime,'' {\em Astrophys. J. Suppl.} {\bf 207}, 6  (2013).

\bibitem{fan2018analytical}
Z.~Fan and H.~Feng, ``The exact solution to null geodesics at the equatorial
  plane in kerr spacetime,'' {\em College Physics} {\bf 37}  (2018).

\bibitem{Gralla_2020}
S.~E. Gralla and A.~Lupsasca, ``Null geodesics of the {K}err exterior,'' {\em
  Phys. Rev. D} {\bf 101}  (2020).

\bibitem{Cie_lik_2023}
A.~Cieślik, E.~Hackmann, and P.~Mach, ``Kerr geodesics in terms of
  {W}eierstrass elliptic functions,'' {\em Phys. Rev. D} {\bf 108}  (2023).

\bibitem{Liu__2024}
Y.~Liu and B.~Sun, ``Analytical solutions of equatorial geodesic motion in
  {K}err spacetime*,'' {\em Chinese Phys. C} {\bf 48}, 045107  (2024).

\bibitem{Chandrasekhar1983mathematical}
S.~Chandrasekhar, {\em The Mathematical Theory of Black Holes}, Oxford
  University Press, Oxford  (1983).

\bibitem{Carter_1968}
B.~Carter, ``Global structure of the kerr family of gravitational fields,''
  {\em Physical Review} {\bf 174}, 1559–1571  (1968).

\bibitem{hagihara1930theory}
Y.~Hagihara, ``Theory of the relativistic trajeetories in a gravitational field
  of {S}chwarzschild,'' {\em Japanese Journal of Astronomy and Geophysics, Vol.
  8, p. 67} {\bf 8}, 67  (1930).

\bibitem{gradshteyn2000table}
I.~S. Gradshteyn and I.~M. Ryzhik, {\em Table of Integrals, Series, and
  Products}, Academic Press, Singapore, 6th~ed.  (2000).

\bibitem{Bret_n_2017}
N.~Bretón, O.~d.~J. Cabrera-Rosas, E.~Espíndola-Ramos, {\em et~al.},
  ``Towards the {R}onchi test for gravitational lenses: the
  gravitoronchigram,'' {\em J. Opt.} {\bf 19}, 065602  (2017).

\bibitem{Gomez-Correa:21}
J.~E. G\'{o}mez-Correa, A.~L. Padilla-Ortiz, A.~Jaimes-N\'{a}jera, {\em
  et~al.}, ``Generalization of ray tracing in symmetric gradient-index media by
  {F}ermat's ray invariants,'' {\em Opt. Express} {\bf 29}, 33009--33026
  (2021).

\bibitem{Lakshminarayanan}
V.~Lakshminarayanan, A.~K. Ghatak, and K.~Thyagarajan, {\em Lagrangian optics},
  Springer  (2002).

\bibitem{Luneburg_1964}
R.~K. Luneburg, {\em Mathematical Theory of Optics}, University of California
  Press  (1964).

\bibitem{Gomez-Correa:22}
J.~E. G\'{o}mez-Correa, ``Geometrical-light-propagation in non-normalized
  symmetric gradient-index media,'' {\em Opt. Express} {\bf 30}, 33896--33910
  (2022).

\bibitem{Born_2019}
M.~Born and E.~Wolf, {\em Principles of Optics: 60th Anniversary Edition},
  Cambridge University Press  (2019).

\bibitem{Nouri_Zonoz_2022}
M.~Nouri-Zonoz, A.~Parvizi, and H.~Forghani-Ramandy, ``Metamaterial analog of a
  black hole shadow: An exact ray-tracing simulation based on the spacetime
  index of refraction,'' {\em PRD} {\bf 106}  (2022).

\bibitem{burden2011numerical}
R.~L. Burden and J.~D. Faires, {\em Numerical Analysis}, Brooks/Cole, Cengage
  Learning, Boston, MA, 9th~ed.  (2011).

\bibitem{goldstein2002classical}
H.~Goldstein, C.~P. Poole, and J.~L. Safko, {\em Classical Mechanics}, Addison
  Wesley, Boston, 3rd~ed.  (2002).

\bibitem{Ashkenazy_2025}
Y.~Ashkenazy and S.~Balberg, ``The s2 orbit and tidally disrupted binaries:
  Indications for collisional depletion in the galactic center,'' {\em
  Astronomy \&amp; Astrophysics} {\bf 695}, A98  (2025).

\bibitem{Xiao_2021}
W.~Xiao, S.~Tao, and H.~Chen, ``Mimicking the gravitational effect with
  gradient index lenses in geometrical optics,'' {\em Photonics Res.} {\bf 9},
  1197  (2021).

\bibitem{Hall_2022}
M.~J.~W. Hall, ``Simple precession calculation for mercury: A linearization
  approach,'' {\em American Journal of Physics} {\bf 90}, 857–860  (2022).

\bibitem{rindler2006relativity}
W.~Rindler, {\em Relativity: special, general, and cosmological}, Oxford
  University Press  (2006).

\bibitem{Lippman_2021}
D.~H. Lippman, N.~S. Kochan, T.~Yang, {\em et~al.}, ``Freeform gradient-index
  media: a new frontier in freeform optics,'' {\em Opt. Express} {\bf 29},
  36997  (2021).

\bibitem{Lemon_2024}
C.~Lemon, F.~Courbin, A.~More, {\em et~al.}, ``Searching for strong
  gravitational lenses,'' {\em Space Sci. Rev.} {\bf 220}  (2024).

\bibitem{Mendoza_2024}
S.~Mendoza and M.~Santibañez-Armenta, ``A comparison between the deflection
  angles of massive and massless particles in the {S}chwarzschild space-time
  and their consequences on black hole shadows,'' {\em IJGMMP} {\bf 21}
  (2024).

\bibitem{Atamurotov_2015}
F.~Atamurotov, B.~Ahmedov, and A.~Abdujabbarov, ``Optical properties of black
  holes in the presence of a plasma: The shadow,'' {\em PRD} {\bf 92}  (2015).

\bibitem{Rogers_2015}
A.~Rogers, ``Frequency-dependent effects of gravitational lensing within
  plasma,'' {\em MNRAS} {\bf 451}, 17–25  (2015).

\bibitem{Matt_1993}
G.~Matt, ``X-ray polarization properties of a centrally illuminated accretion
  disc,'' {\em MNRAS} {\bf 260}, 663–674  (1993).

\bibitem{Beloborodov_1998}
A.~M. Beloborodov, ``Polarization change due to fast winds from accretion
  disks,'' {\em The Astrophysical Journal} {\bf 496}, L105–L108  (1998).

\bibitem{Isaacson1968}
R.~A. Isaacson, ``Gravitational radiation in the limit of high frequency. i.
  the linear approximation and geometrical optics,'' {\em Physical Review} {\bf
  166}(5), 1263--1271  (1968).

\bibitem{Berry_1980}
M.~Berry and C.~Upstill, {\em IV Catastrophe Optics: Morphologies of Caustics
  and Their Diffraction Patterns}, 257–346.
\newblock Elsevier  (1980).

\bibitem{Cusin_2020}
G.~Cusin and M.~Lagos, ``Gravitational wave propagation beyond geometric
  optics,'' {\em PRD} {\bf 101}  (2020).

\end{thebibliography}
\bibliographystyle{spiejour}   

\end{document}